\documentclass[%
 reprint,
 amsmath,amssymb,
 aps,
floatfix,
]{revtex4-2}

\pdfoutput=1

\usepackage{textcomp}
\usepackage{natbib}
\usepackage{graphicx}
\usepackage{siunitx}
\usepackage{xcolor}  

\begin{document}

\title{Morphology-induced spectral modification of self-assembled WS$_2$ pyramids}

\author{Irina Komen}
\author{Sabrya E. van Heijst}
\author{Sonia Conesa-Boj}
\author{L. Kuipers\thanks{L.Kuipers@tudelft.nl}}
\date{\today}

\affiliation{Kavli Institute of Nanoscience, Department of Quantum Nanoscience, Delft University of Technology, The Netherlands}

\begin{abstract}
Due to their intriguing optical properties, including stable and chiral excitons, two-dimensional transition metal dichalcogenides (2D-TMDs) hold the promise of applications in nanophotonics. Chemical vapor deposition (CVD) techniques offer a platform to fabricate and design nanostructures with diverse geometries. However, the more exotic the grown nanogeometry, the less is known about its optical response. WS$_2$ nanostructures with geometries ranging from monolayers to hollow pyramids have been created. The hollow pyramids exhibit a strongly reduced photoluminescence with respect to horizontally layered tungsten disulphide, facilitating the study of their clear Raman signal in more detail. Excited resonantly, the hollow pyramids exhibit a great number of higher-order phononic resonances. In contrast to monolayers, the spectral features of the optical response of the pyramids are position dependent. Differences in peak intensity, peak ratio and spectral peak positions reveal local variations in the atomic arrangement of the hollow pyramids crater and sides. The position-dependent optical response of hollow WS$_2$ pyramids is characterized and attributed to growth-induced nanogeometry. Thereby the first steps are taken towards producing tunable nanophotonic devices with applications ranging from opto-electronics to non-linear optics.
\end{abstract}

\maketitle

\section{Introduction}

Two-dimensional transition metal dichalcogenites (2D-TMDs) have recently attracted enormous scientific attention for their distinctive optical properties. Like graphene, TMDs materials consist of layers held together only by van der Waals forces. 2D TMDs (\textit{e.g.} MoS$_2$, WS$_2$, MoSe$_2$, WSe$_2$ \textit{etc.}) are atomically thin semiconductors, with a transition from indirect to direct bandgap in the few-layer limit \cite{Mak_MoS2monofirst_PhysRevLett_2010}. Due to their high binding energy, electron-hole pairs form stable excitons even at room-temperature. A pseudospin can be attributed to each of the two valleys 2D TMDs possess, making it possible to address them selectively with circularly polarized light \cite{Cao_TMDCcircular_NatCom_2012, Xu_TMDCspins_NatPhys_2014, Zhu_WS2bilayerValleyPolarization_PNAS_2014}. The electronic and optical properties of these 2D semiconductors, for instance the excitons and valley pseudospin, make them an interesting platform for opto-electronics \cite{Zhang_TMDCtransistor_science_2014, Wang_TMDCelectronics_NatNano_2012} and valleytronics \cite{Mak_TMDCvalleyHall_science_2014, Schaibley_valleytronics_NatRev_2016, Irina_2020} applications.

One of the ways of creating 2D TMD monolayer samples is by exfoliation from bulk \cite{Irina_2020}. An alternative method to produce layered TMDs materials is Chemical Vapor Deposition (CVD) \cite{Song_CVDgrownWS2_ACSNano_2013, Zhang_CVDgrownWS2_ACSNano_2013, Cong_CVDgrownWS2_AdvOptMat_2014, Orofeo_CVDgrownWS2_APL_2014, Thangaraja_WS2crystals_MatLett_2015, Liu_CVDgrownWS2_NanoscResLett_2017}. While CVD can reproduce horizontal layers as found in naturally occurring TMDs, adjusting the growth conditions enables the growth of nanostructures with exciting properties and applications in nanotechnology, for instance vertical walls, flowers and pyramids \cite{Sabrya2020}.
Under certain CVD growth parameters TMDs materials form pyramid-like structures \cite{Zhang_MoS2spirals_NanoLett_2014, Sarma_spiralWS2_RSCAdv_2016}, having many active adsorption site useful for applications in hydrogen sensors \cite{Agrawal_MoS2pyramids_JEnergy_2020} and water disinfection \cite{Cheng_MoS2pyramids_MatInterf_2018}, at the same time possessing interesting electronic properties like ferromagnetism \cite{Zhou_MoS2pyramid_Nanoscale_2018} and high mobility \cite{Zheng_MoS2pyramids_AdvMat_2017, Chen_WSe2pyramids_ACSNano_2014}. Moreover, pyramid-like TMDs structures have applications in non-linear optics \cite{Fan_SHGpyramid_ACSNano_2017, Zheng_MoS2pyramids_AdvMat_2017, Lin_SHGpyramids_ACSNano_2018}, as they exhibit higher non-linear optical conversion efficiency than monolayers due to the thickness increase, while demonstrating a much larger non-linear optical response than multilayer TMDs. 

It is interesting to note that only in a limited number of studies in the literature photoluminescence of these spiral or pyramid like structures is reported \cite{Zhang_MoS2spirals_NanoLett_2014, Sarma_spiralWS2_RSCAdv_2016, Zheng_MoS2pyramids_AdvMat_2017,  Cheng_MoS2pyramids_MatInterf_2018}. In some, the non linear optical response is studied \cite{Fan_SHGpyramid_ACSNano_2017, Zheng_MoS2pyramids_AdvMat_2017, Lin_SHGpyramids_ACSNano_2018}. In most, the TMDs spirals and pyramids are studied using Raman spectroscopy \cite{Chen_WSe2pyramids_ACSNano_2014, Zhang_MoS2spirals_NanoLett_2014,  Sarma_spiralWS2_RSCAdv_2016,  Zheng_MoS2pyramids_AdvMat_2017, Fan_SHGpyramid_ACSNano_2017,  Zhou_MoS2pyramid_Nanoscale_2018, Cheng_MoS2pyramids_MatInterf_2018,  Agrawal_MoS2pyramids_JEnergy_2020}. However, comparing the measured optical response of pyramid-like structures of different studies needs to be done with caution, as the terms spiral flake or pyramid are used for nanostructures that have different thicknesses, geometry and sizes. 

Raman spectroscopy is a powerful tool to study 2D TMDs, as knowledge on the vibrational modes of the layers provides insights in their structure  \cite{Lee_MoS2Ramanfirst_ACSNano_2010,  Zhao_RamanTMDlinear_Nanoscale_2013, Berkdemir_RamanWS2_ScientRep_2013}. Commonly studied are the characteristic vibrational modes of TMDs, the E$^1_{2g}$ that corresponds to the in-plane displacement of the atoms, and the A$_{1g}$ that corresponds to the out-of-plane displacement of the chalcogenide atoms, as well as the longitudinal acoustic phonon LA(M) \cite{Berkdemir_RamanWS2_ScientRep_2013,Zhao_RamanTMDlinear_Nanoscale_2013,  Mitioglu_RamanWS2LAmode_PRB_2014, Molas_RamanWS2_ScientRep_2017}. Studying the Raman response as a function of temperature \cite{Peimyoo_temperatureRamanWS2_NanoRes_2015, Gaur_temperatureRamanWS2_PhysChemC_2015, Li_MoS2Raman_pressureshift_APL_2016, Li_RamanWS2_defectsT_JPhysChemC_2019, Fan_resonanceRamanTMD_JApplPhys_2014}, excitation polarization \cite{Mitioglu_RamanWS2LAmode_PRB_2014, Zhao_RamanTMDlinear_Nanoscale_2013, Chen_helicityRamanTMD_NanoLett_2015} and excitation wavelength \cite{Carvalho2015, Corro_resonantRamanTMD_NanoLetters_2016, McDonnell_resonantRamanWS2_NanoLetters_2018}, provides information on structural properties like number of layers, strain and defect density. 

It is important to note that the existence of exciton resonances in 2D TMDs has large implications for its Raman response. Raman features are greatly enhanced when the excitation is in resonance with an excitonic transition \cite{Zucker_resonantExciton_PhysRevLett_1983, Berkdemir_RamanWS2_ScientRep_2013,Zhao_RamanTMDlinear_Nanoscale_2013, Carvalho2015, Corro_resonantRamanTMD_NanoLetters_2016, McDonnell_resonantRamanWS2_NanoLetters_2018}. Resonance Raman spectroscopy on 2D TMDs materials results in the excitation of higher-order phononic resonances \cite{Golasa_multiphononMoS2_APL_2014}, which yields rich Raman spectra with many more modes than the two mentioned characteristic modes. 

Many questions arise about the nature of the optical response from complex CVD-grown TMDs pyramid like nanostructures. It is unknown how the nanogeometry of TMDs pyramids influences its photoluminescence and Raman response and how it depends on temperature and polarization. For potential applications, this knowledge is paramount. 

\begin{figure*}[htp]
\centering
\includegraphics[width = 0.8\linewidth] {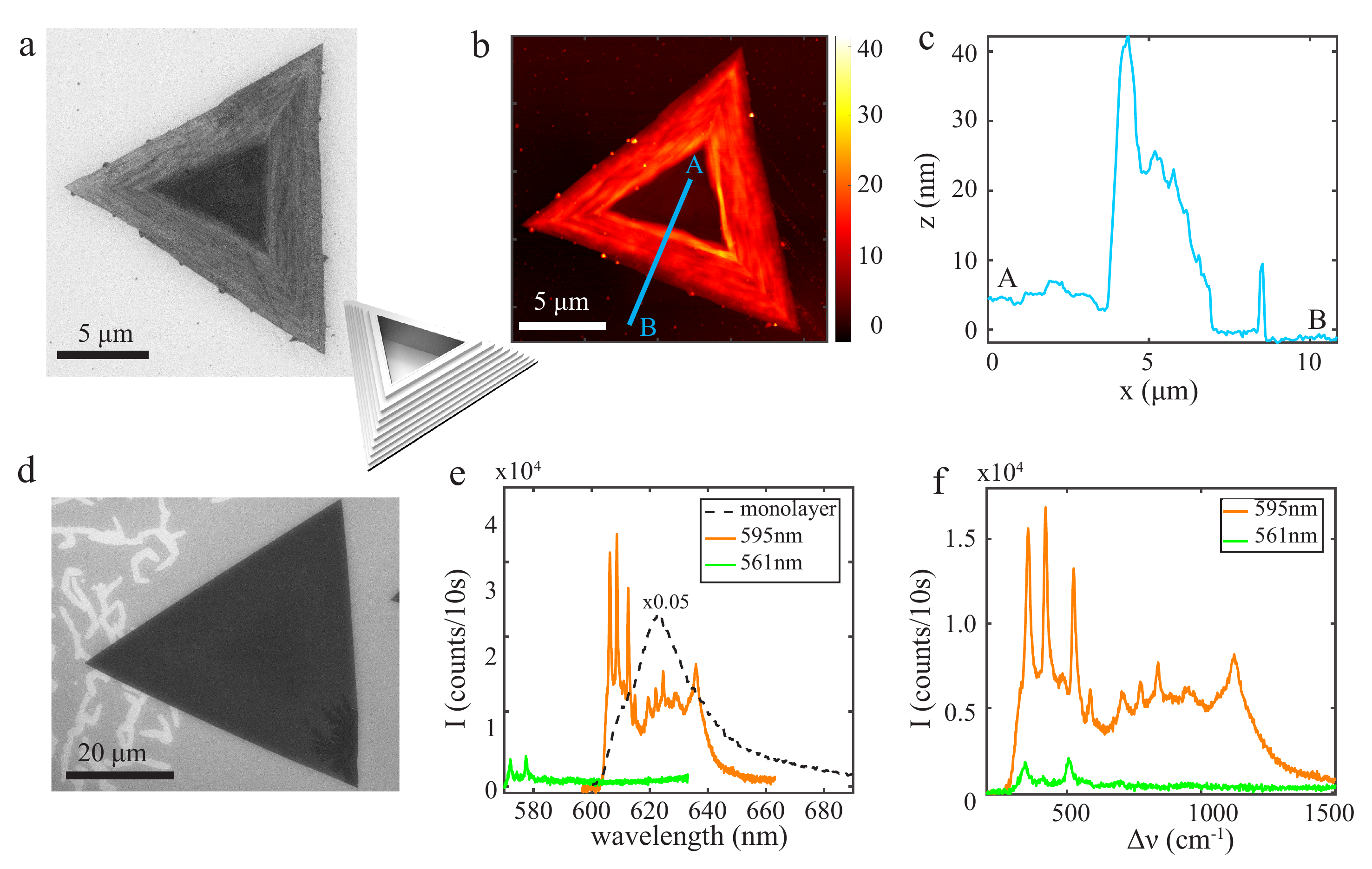}
\caption{\textcolor{black}{\textbf{Hollow WS$_2$ pyramids}} \\ \textcolor{black}{\textbf{a.} SEM image of the hollow WS$_2$ pyramid. The lines along the sides indicate the single stair steps, whereas the darker region in the middle is the crater. \textit{inset.} Schematic representation of the hollow pyramid. \textbf{b.} AFM image of the hollow WS$_2$ pyramid. The blue line indicates the position of the AFM crosscut depicted in \textbf{c}, ranging from \textbf{A} the pyramid crater to \textbf{B} the substrate. \textbf{d.} SEM image of a WS$_2$ monolayer on the same substrate. \textbf{e.} Photoluminescence spectrum of the WS$_2$ monolayer (black dotted line), and spectra of a pyramid obtained with a \SI{595}{nm} excitation (orange) and a \SI{561}{nm} excitation (green). When converting the x-axis to wavenumber $\Delta\nu$ in \textbf{f}, the spectral response on the two different lasers overlaps nicely, indicating that the collected light originates from Raman processes rather than photoluminescence.}}
\label{fig_intro}
\end{figure*}

Here, we study the Raman and photoluminesence response of CVD-grown hollow WS$_2$ pyramids, comparing it to the optical response of WS$_2$ monolayers. Even though the WS$_2$ monolayers and pyramids are grown on the same substrate and under the same conditions, their measured optical response is completely different. We find, surprisingly, that the pyramids exhibit a strongly reduced photoluminescence (PL) with respect to horizontal layers. The reduced PL enables us to study the Raman signal of the hollow WS$_2$ pyramids, that contains both the characteristic Raman peaks of flat layers and a great number of higher-order phononic resonances. In contrast with the monolayers, the measured optical response of the hollow WS$_2$ pyramids is non-uniform over the nanostructures. Annular dark-field (ADF) scanning transmission electron microscopy (STEM) measurements confirm position-dependent variations in atomic arrangement.

\section{Results and discussion}

\subsection{Hollow WS$_2$ pyramids}

\textbf{Figure \ref{fig_intro}a} presents an SEM image of a CVD-grown hollow WS$_2$ pyramid (the substrate is a SiN film on Si, see Methods). The WS$_2$ is crystallized in a 3R-phase (see Fig.\ref{fig_TEM-morp-and-3R}c-d and Section A in the Supplementary Materials.) The clear lines along the pyramid sides indicate single steps (see Fig.\ref{fig_microscope}c and Fig.\ref{fig_TEM-morp-and-3R}a,b in the Supplementary Materials). The geometry of the darker middle becomes more clear when examining the AFM image in Fig.\ref{fig_intro}b. The height profile measured with the AFM along the blue line is presented in Fig.\ref{fig_intro}c. The bottom of the crater in the middle is roughly \SI{5.6}{nm} high with respect to the substrate, whereas the pyramid sides reach a height of \SI{44}{nm}. The inset of Fig.\ref{fig_intro}a displays a schematic representation of the hollow pyramid, depicting the stair-like sides in white and the crater with a bottom of finite thickness in the middle.

\mbox{Figure \ref{fig_intro}e} presents optical spectra obtained by exciting a WS$_2$ pyramid \textcolor{black}{at the stair-like side} with either a \SI{595}{nm} laser (orange) or a \SI{561}{nm} laser (green). The spectra contain a similar sequence of peaks, but their spectral position does not overlap in wavelength. However, these peaks are located at the same relative frequency distance to the excitation laser, as depicted in Fig.\ref{fig_intro}f. The clear overlap of the larger peak positions in the spectra from the two different lasers indicates that the collected light originates from an inelastic Raman process rather than from photoluminescence. 

\mbox{Figure \ref{fig_intro}d} presents an SEM image of a single horizontal layer (monolayer) WS$_2$ grown on the same substrate (next to hollow pyramids and monolayers, also fully grown pyramids are present on this substrate, see Supplementary Materials, Fig.\ref{fig_microscope}a, Fig.\ref{fig_ratio} and Fig.\ref{fig_peak_pos}). Comparing in Fig.\ref{fig_intro}e the optical response of the monolayer (black dotted line) with that of the pyramid, it becomes apparent that the pyramid exhibits a strongly reduced photoluminescence with respect to monolayer WS$_2$. A small background under the Raman modes is visible in the spectrum of the pyramid around the PL wavelength of \SI{630}{nm}. If this is remnant PL emerging from the hollow WS$_2$ pyramid, it has an intensity of at most 1\% of the WS$_2$ monolayer (see Supplementary Materials Fig.\ref{fig_background}). The immense reduction of the photoluminescence intensity from the hollow pyramids is unexpected. Even though the pyramid crater is only a few nanometers thin, the pyramid spectra do not resemble spectra of standard few-layered horizontal WS$_2$ (see Supplementary Materials Fig.\ref{fig_background}f). When the WS$_2$ thickness increases from monolayer to bulk, it transitions from a direct to an indirect bandgap semiconductor \cite{Mak_MoS2monofirst_PhysRevLett_2010}. It is important to note that, in contrast to the hollow pyramids, the reduced PL from the direct bandgap can be easily measured for few-layered WS$_2$ \cite{Irina_2020, Mak_MoS2monofirst_PhysRevLett_2010}. Moreover, the hollow WS$_2$ pyramids do not exhibit any luminescence at the known wavelength of the indirect bandgap, as would have been expected from both bulk and few-layer WS$_2$ (see Supplementary Materials Fig.\ref{fig_background}f). \textcolor{black}{Therefore we conclude that the increase of the layer number from monolayer WS$_2$ to the pyramid crater cannot explain the reduction of the PL intensity.} CVD-grown horizontal 2D TMDs flakes exhibit a similar amount of photoluminescence as exfoliated samples. Hence, the PL intensity reduction by at least two orders of magnitude of the hollow pyramids cannot merely be explained as being intrinsic to the CVD growth process, \textit{e.g.} through an increase in defect density. Furthermore, the 3R-WS$_2$ nanostructure cannot explain the PL intensity reductions, as 3R-WS$_2$ exhibits the same photoluminescence as the naturally occurring 2H-WS$_2$ \cite{Yang_WS2RamanPL3R_Nanotech_2019, Du_3RcircularPL_PRB_2019}.

We conclude that our hollow WS$_2$ pyramids have a lower quantum efficiency than WS$_2$ flakes: assuming that the optical absorption of a WS$_2$ monolayer and a pyramid is the same, the quantum efficiency of these pyramids is at least two orders of magnitude lower than that of a monolayer WS$_2$. Given the fact that the pyramids have a thickness of \SI{5}{nm} - \SI{44}{nm}, it is safe to assume that a pyramid actually absorbs more than a monolayer, therefore the quantum efficiency is likely to be at least another order of magnitude lower. We attribute the decrease in the quantum efficiency to the increase in possible non-radiative loss channels due to the presence of all the edges in the structure of these pyramids. \textcolor{black}{The increase in non-radiative loss channels due to the edges in the pyramid structure is therefore the main factor that} leads to a severe quenching of the exciton photoluminescence, without influencing the Raman modes. 

The reduction of the PL enables us to study the Raman response of the hollow WS$_2$ pyramids in more detail, as many Raman features are usually obscured by the PL spectrum. The pyramid spectra obtained with the \SI{595}{nm} excitation exhibit roughly 10-12 Raman features, three of which have not been reported before for neither horizontal WS$_2$ layers nor WS$_2$ nanostructures (see Fig.\ref{fig_determine}). For the \SI{561}{nm} excitation, fewer Raman features are visible in the spectra, and these features have a lower intensity. This can be attributed to the fact that the \SI{595}{nm} excitation light is close to the A-exciton resonance, whereas the \SI{561}{nm} is out-of-resonance with both the A- and B-excitons. Raman modes of TMDs can be greatly enhanced when they are excited in resonance with an excitonic transition \cite{Zucker_resonantExciton_PhysRevLett_1983, Berkdemir_RamanWS2_ScientRep_2013,Zhao_RamanTMDlinear_Nanoscale_2013,  Corro_resonantRamanTMD_NanoLetters_2016, McDonnell_resonantRamanWS2_NanoLetters_2018}.

\subsection{\textcolor{black}{Structural characterization}}

\begin{figure*}[htp]
\centering
\includegraphics[width = \linewidth] {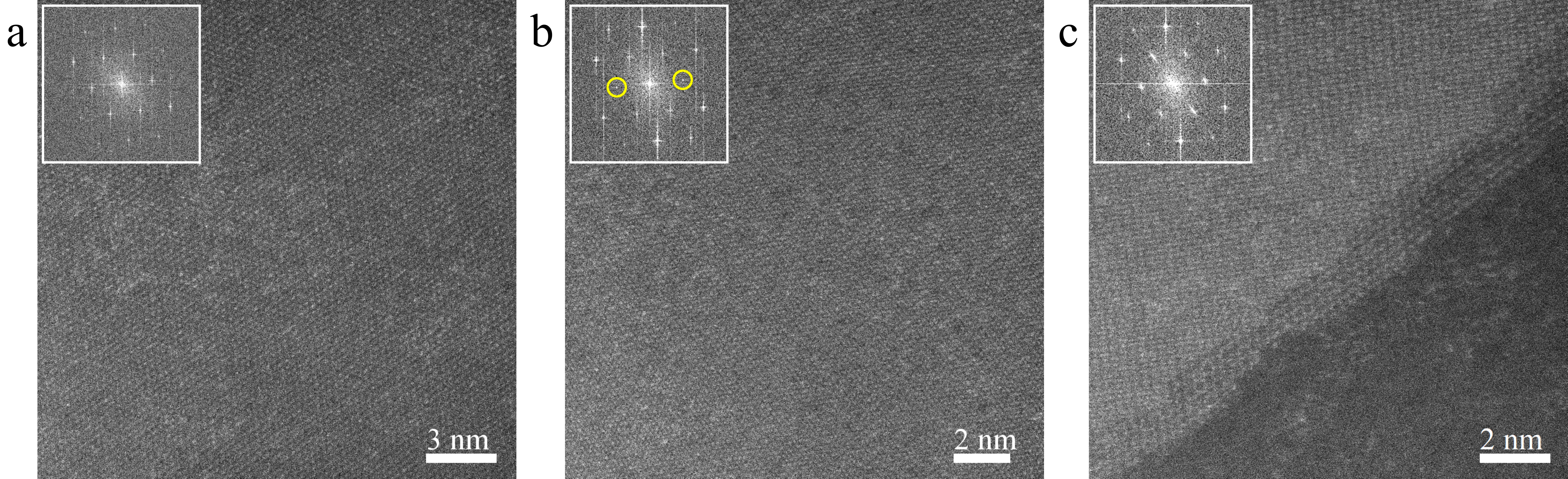}
\caption{\textcolor{black}{\textbf{ADF-STEM of a hollow WS$_2$ pyramid}} \\ Atomic resolution annular dark-field scanning transmission electron microscopy images taken in \textbf{a-b} the middle region, and \textbf{c.} the side of a hollow WS$_2$ pyramid. Subtle variations in the atomic arrangement can be observed. The corresponding FFTs, as given as insets in each of the panels, highlight this feature further. Changes in the Bragg reflections occur between them. One of which is a reduction of the intensity of one of the Bragg reflections in the middle region of the hollow pyramid, as marked by the yellow circles in \textbf{b}.}
\label{fig_TEM_mt}
\end{figure*}

In order to help interpret the \textcolor{black}{studied} optical response, we perform a detailed structural characterization of the hollow WS$_2$ pyramids by means of Transmission Electron Microscopy measurements. \textbf{Figure \ref{fig_TEM_mt}} displays annular dark-field (ADF) scanning transmission electron microscopy (STEM) images taken in different regions of these WS$_2$ nanostructures. Figures \ref{fig_TEM_mt}a and \ref{fig_TEM_mt}b display atomic resolution ADF-STEM images of two different locations corresponding to the middle region of the hollow pyramid, while Fig.\ref{fig_TEM_mt}c corresponds to the pyramid side. By comparing these three images, we can clearly observe differences in the atomic arrangement. In Figure \ref{fig_TEM_mt}a, the atomic distribution displays a well defined hexagonal shape, and this symmetry is also highlighted by the corresponding fast Fourier transform (see inset in Fig.\ref{fig_TEM_mt}a). As we can observe from Fig.\ref{fig_TEM_mt}b, the atomic arrangements at this location are somewhat different from those observed in Fig.\ref{fig_TEM_mt}a. This structural variation is also reflected in the corresponding FFT (inset in Fig.\ref{fig_TEM_mt}b), where one of the Bragg reflections exhibits a reduced intensity (marked with a yellow circles to facilitate its visualization). Structural variation is also observed in the side region of the hollow pyramid. The difference in contrast in Fig.\ref{fig_TEM_mt}c corresponds to two stair-like steps in the pyramid side. By comparing their relative atomic arrangement, we can determine that, while the external step exhibits a clear hexagonal honeycomb structure, this arrangement is lost in the subsequent layer. Note also that, as observed from the results obtained in the middle region of the nanostructures, even across its sides the atomic arrangement can vary slightly (see also Figure \ref{fig_additional_TEM} in the Supplementary Materials).

These subtle variations of the atomic arrangement might be induced by the local presence of strain, which in turn results into a slight change of the orientation of the flake. Importantly, the level of structural disorder is more marked in the middle of the pyramids as compared to the sides (see Figure \ref{fig_additional_TEM}) due to the additional presence of free-standing WS$_2$ flakes arising from the walls of the hollow pyramid.

\begin{figure*}[htp]
\centering
\includegraphics[width = 0.6\linewidth] {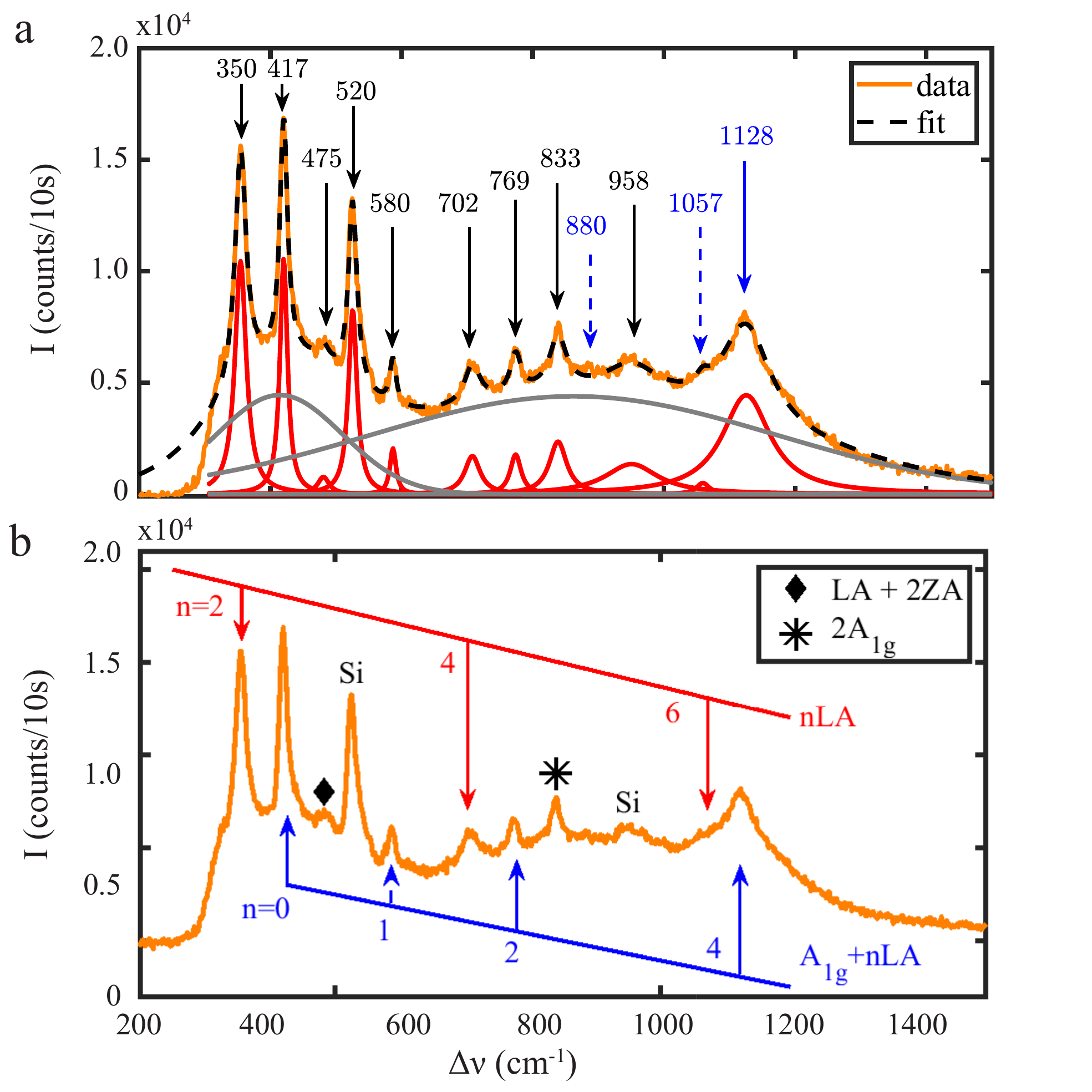}
\caption{\textbf{Characterization of Raman peaks} \\ \textcolor{black}{\textbf{a.} The optical response of the hollow pyramids (orange) can be fitted by eleven Lorentzian lineshapes (red) and two Gaussians (grey). The Raman features are indicated by arrows with their spectral position in cm$^{-1}$. The last three features (in blue) have not been reported before. \textbf{b.} Part of the features can be explained as being multiphonon resonances involving the LA(M) phonon. The blue line depicts the higher order resonances of $A_{1g}$+n*LA(M). The red line depicts the higher order resonances of n*LA(M). }}
\label{fig_determine}
\end{figure*}

\subsection{Characterization of vibrational modes}

In \textbf{Figure \ref{fig_determine}a} we present a WS$_2$ pyramid spectrum\textcolor{black}{, acquired at the pyramid side,} in which all Raman features are indicated with arrows. Commonly, only three Raman modes are measured on both horizontal TMDs layers or nanostructures. We measure 10-12 Raman features, three of which have not been reported previously (indicated in blue in Fig.\ref{fig_determine}a). The other modes can be attributed following a limited amount of previous investigations (see also Table \ref{table_peaks}).
In order to analyze the spectra in more detail, we fit the overall spectrum with a collection of eleven Lorentzian lineshapes (red) and a background consisting of two Gaussians (grey) (see Supplementary Materials Section 1 for a discussion on the background). This way we are able to attribute the new Raman modes from hollow WS$_2$ pyramids to being multiphonon resonances involving the LA(M) phonon, adopting the methodology for high frequency Raman features in  MoS$_2$ \cite{Golasa_multiphononMoS2_APL_2014}. The blue line in Fig.\ref{fig_determine}b depicts the higher order resonances of $A_{1g}$+n*LA(M). The peak \SI{580}{cm^{-1}} is commonly attributed to $A_{1g}$+LA(M) \cite{Molas_RamanWS2_ScientRep_2017, Berkdemir_RamanWS2_ScientRep_2013, Peimyoo_temperatureRamanWS2_NanoRes_2015, Gaur_temperatureRamanWS2_PhysChemC_2015}, and the peak at \SI{769}{cm^{-1}} is attributed to $A_{1g}$+2LA(M) by Molas \textit{et al} \cite{Molas_RamanWS2_ScientRep_2017}. Thus we attribute the newly observed peak at \SI{1128}{cm^{-1}} (n=4) to $A_{1g}$+4LA(M). 
The red line in Fig.\ref{fig_determine}b depicts the higher order resonances of n*LA(M). The peak at \SI{702}{cm^{-1}} is commonly attributed to 4LA(M) \cite{Berkdemir_RamanWS2_ScientRep_2013, Peimyoo_temperatureRamanWS2_NanoRes_2015}, which is twice the first 2LA(M) Raman peak at \SI{350}{cm^{-1}}. Therefore we attribute the newly observed small shoulder of the last peak around \SI{1057}{cm^{-1}} to 6LA(M).
The expected resonance at 3LA(M) (red line in Fig.\ref{fig_determine}b) would spectrally overlap with the first silicon peak at \SI{520}{cm^{-1}}, as well as the expected resonance at n=3 (blue line in Fig.\ref{fig_determine}b) would overlap with the second silicon peak \SI{955}{cm^{-1}}, so these possible features cannot be distinguished from the substrate response. The expected 5LA(M) (red line in Fig.\ref{fig_determine}b) would be around \SI{880}{cm^{-1}}, but is too dim to distinguish very clearly from the background. The peak at \SI{833}{cm^{-1}} is 2*$A_{1g}$, and the peak at \SI{475}{cm^{-1}} is commonly attributed to LA(M)+2ZA(M) (see Table \ref{table_peaks}). 
We conclude that the observed high-frequency Raman modes are multiphonon resonances involving the LA(M) phonon, excited because the \SI{595}{nm} laser is in resonance with the A-exciton. 

The highest frequency Raman modes have not been reported before on horizontal WS$_2$ layers (indicated in blue in Fig.\ref{fig_determine}a). A possible explanation for not observing them on monolayers is that investigating Raman modes on horizontal WS$_2$ layers is experimentally challenging because of the presence of photoluminescence, that is much brighter than any Raman feature. An intriguing alternative hypothesis is that the nanogeometry of the hollow pyramid plays a role in exciting the higher order Raman modes more resonantly, \textit{e.g.} through a higher phonon density of states. 

\begin{table*}[htp]
\begin{center}
  \begin{tabular}{| c | c | c | c | c | c |}
    \hline
    position & std (cm$^{-1}$) & brightness & attributed to & possibly & literature \\ \hline
    350 cm$^{-1}$ &  1.6  & ++ & E$^1_{2g}$ / 2LA(M) & & \cite{Berkdemir_RamanWS2_ScientRep_2013,Zhao_RamanTMDlinear_Nanoscale_2013,  Mitioglu_RamanWS2LAmode_PRB_2014,  Corro_resonantRamanTMD_NanoLetters_2016, Peimyoo_temperatureRamanWS2_NanoRes_2015, Thripuranthaka_temperatureRamanTMD_APL_2014, Molas_RamanWS2_ScientRep_2017} \\ \hline
    417 cm$^{-1}$ & 1.5  & ++ & A$_{1g}$ & & \cite{Berkdemir_RamanWS2_ScientRep_2013,Zhao_RamanTMDlinear_Nanoscale_2013,  Corro_resonantRamanTMD_NanoLetters_2016, Peimyoo_temperatureRamanWS2_NanoRes_2015, Thripuranthaka_temperatureRamanTMD_APL_2014, Molas_RamanWS2_ScientRep_2017} \\
    \hline
    475 cm$^{-1}$ & 4.7  & - - & LA + 2ZA  & &  \cite{Molas_RamanWS2_ScientRep_2017} \\
     & & & or E''(M) + TA(M) & & \cite{McDonnell_resonantRamanWS2_NanoLetters_2018} \\
    \hline
    520 cm$^{-1}$ & 0.8  & ++ & Si & 3LA(M) & \cite{Parker_Silicon_PhysRev_1967, Uchinokura_Silicon_SolStateComm_1972, Weinstein_Silicon_SolStateComm_1972} \\
    \hline
    580 cm$^{-1}$ & 1.0  & +- & A$_{1g}$ + LA(M) & & \cite{Berkdemir_RamanWS2_ScientRep_2013, Peimyoo_temperatureRamanWS2_NanoRes_2015, Thripuranthaka_temperatureRamanTMD_APL_2014,Molas_RamanWS2_ScientRep_2017,Gaur_temperatureRamanWS2_PhysChemC_2015} \\
    \hline
    702 cm$^{-1}$ & 1.4  & + & 4LA(M) & 2E$^1_{2g}$ & \cite{Berkdemir_RamanWS2_ScientRep_2013, Peimyoo_temperatureRamanWS2_NanoRes_2015, Thripuranthaka_temperatureRamanTMD_APL_2014} (\cite{Molas_RamanWS2_ScientRep_2017}) \\
    \hline
    769 cm$^{-1}$ & 1.0  & + & A$_{1g}$ + E$^1_{2g}$ & A$_{1g}$ + 2LA(M) & \cite{Molas_RamanWS2_ScientRep_2017} \\
    \hline
    833 cm$^{-1}$ & 1.2  & + & 2A$_{1g}$ & & \cite{Molas_RamanWS2_ScientRep_2017} \\
    \hline
    880 cm$^{-1}$ &  & - - &  & 5LA(M) & \\
    \hline
    958 cm$^{-1}$ & 3.3  & +- & Si & A$_{1g}$ + 3LA(M) & \cite{Parker_Silicon_PhysRev_1967, Uchinokura_Silicon_SolStateComm_1972, Weinstein_Silicon_SolStateComm_1972} \\
    \hline
    1057 cm$^{-1}$ & 1.0  & - - & & 6LA(M) & \\
    \hline
    1128 cm$^{-1}$ & 4.0  & + & & A$_{1g}$ + 4LA(M) & \\
    \hline
  \end{tabular}
\end{center}
\caption{\textcolor{black}{Overview of the measured Raman features on the hollow pyramid (see Fig.\ref{fig_determine}) at room-temperature. The first column gives the peak position determined by fitting, with the statistical standard deviation in the second column and the brightness in the third. The fourth column indicates known Raman modes, where the last column gives a few references. The fifth columns indicates possible explanations, either taken from only one article or being our hypothesis.} }
\label{table_peaks}
\end{table*}

\begin{figure*}[htp]
\centering
\includegraphics[width = 0.9\linewidth] {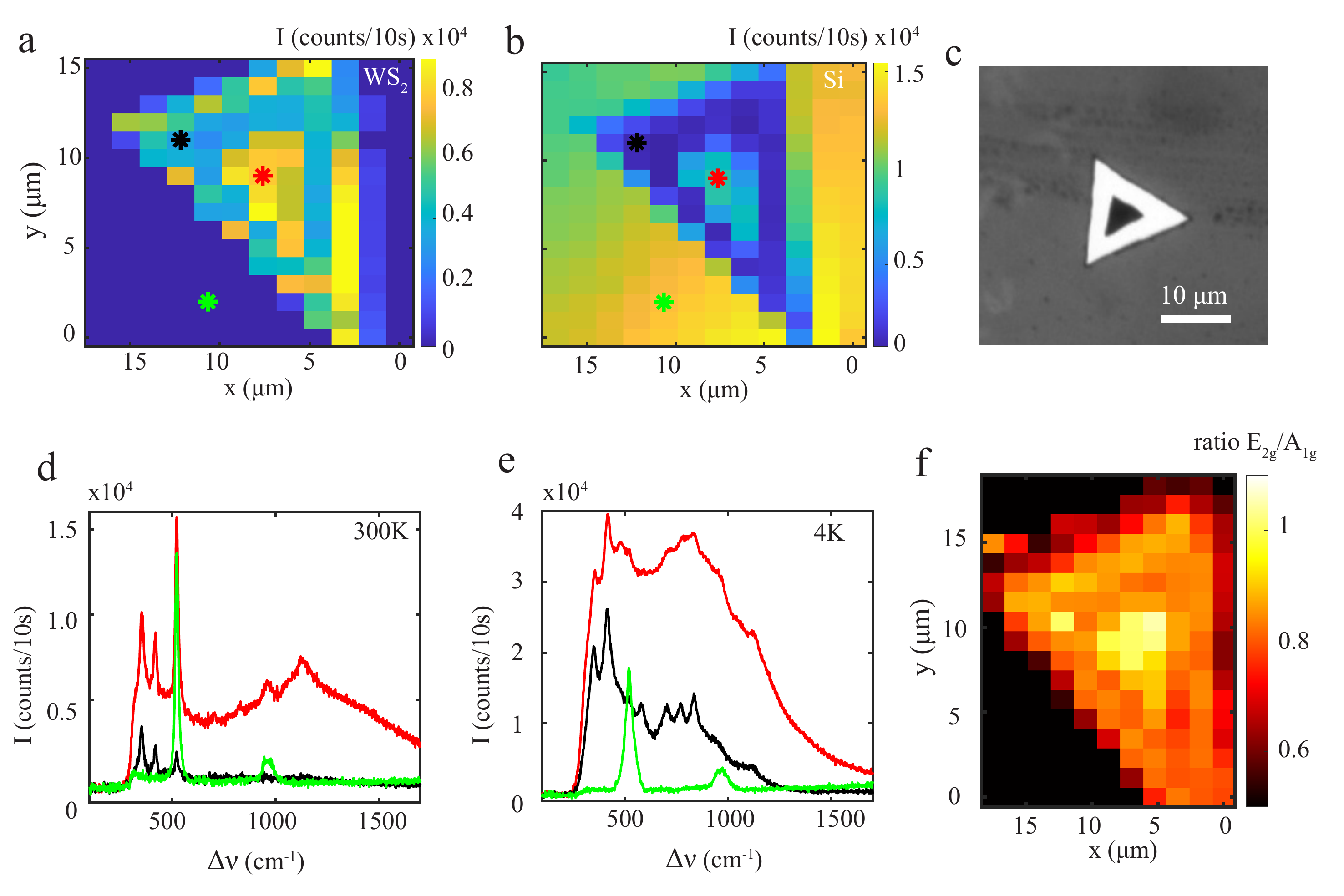}
\caption{\textbf{Position dependence of intensity and shape of spectra} \\ \textcolor{black}{\textbf{a.} Intensity map of the first WS$_2$ Raman feature around \SI{350}{cm^{-1}}, with a step size of around \SI{1.5}{\mu m}. The stars indicate the positions of the spectra in \textbf{d-e}. Note that the x and y axis in the map are slightly skewed due to experimental constraints (see Experimental Section). \textbf{b.} Intensity map of the Si Raman peak at \SI{520}{cm^{-1}}. \textbf{c.} Optical image of a hollow WS$_2$ pyramid. Note the white colour at the sides, indicating a clear increase in scattering from the sides with respect to the top. \textbf{d,e.} Pyramid spectra at \SI{300}{K} and \SI{4}{K}. The substrate spectrum (green) shows the two Si peaks at 520 and \SI{955}{cm^{-1}}. The spectrum at the hollow part of the pyramid (red) has an overall higher intensity than at the side (black). \textbf{f.} Map of the ratio between the first two Raman features in the spectra (E$_{2g}$/A$_{1g}$). }}
\label{fig_position}
\end{figure*}

\subsection{Position dependence of spectral features}

The hollow WS$_2$ pyramids contain two distinct regions: the crater in the middle and the stair-like sides. We find that these regions exhibit a different spectral response. This is in contrast from the more homogeneous spectral response for horizontal WS$_2$ flakes (see Supplementary Materials, Fig.\ref{fig_ratio} and Fig.\ref{fig_peak_pos}). \textbf{Figure \ref{fig_position}a} presents an intensity map of the first Raman peak (E$_{2g}$,2LA). To create this map, the maximum value of the fitted peak is used (see Fig.\ref{fig_determine}a). The peak intensity is higher and the peaks are more pronounced at the pyramid crater than at the stair-like sides. For the other WS$_2$ Raman peaks, as well as at different temperatures and using different excitation wavelengths, this intensity distribution looks similar (see Supplementary Materials Fig.\ref{fig_intensity}). 

\mbox{Figure \ref{fig_position}b} presents an intensity map for the silicon peak at \SI{520}{cm^{-1}}. As expected, there is a constant intensity for the substrate next to the pyramid. It is interesting to note that the intensity of this substrate peak also decreases on the pyramid edges. We hypothesise that light scatters a lot from the stair-like pyramid sides, as seen from the bright white colour of the sides in the optical image (Fig.\ref{fig_position}c). This both reduces the available excitation light to excite Raman modes, and scattering of the resulting Raman response reduces the detected light, including the Raman response of the silicon substrate. 
\mbox{Figures \ref{fig_position}d-e} depict spectra on three different positions: on the hollow pyramid middle (red), on the substrate (green) and on the pyramid side (black) (indicated by stars in Fig.\ref{fig_position}a,b). In both the room temperature spectra in Fig.\ref{fig_position}d and \SI{4}{K} spectra in Fig.\ref{fig_position}e (temperature dependence will be discussed later), the two silicon peaks at 520 and \SI{955}{cm^{-1}} can be clearly distinguished (in green). The spectrum at the pyramid crater (red) has clearly a higher overall intensity. In addition to the Raman peaks, there is also a background visible. Especially at \SI{4}{K}, this background signal becomes extremely high, turning the signal from all the higher frequency Raman modes into mere shoulders. Based on its spectral position, we attribute this background to intermediate gap states or defect states (see Supplementary materials Fig.\ref{fig_background}). It is interesting to note that this background is significantly higher in the pyramid crater than on the sides, which might be originated by the presence of crystallographic defects.

\mbox{Figure \ref{fig_position}f} presents a map of the intensity ratio between the first two Raman peaks (E$_{2g}$/A$_{1g}$). Just like the intensity of the individual peaks, this intensity ratio is also non-uniform over the WS$_2$ pyramid. In the pyramid crater, the peak ratio is approximately 1.0, whereas on the stair-like sides, the A$_{1g}$ peak is higher. The difference in peak ratio between the pyramid crater and sides is also present and much higher for the \SI{561}{nm} excitation (see Supplementary Materials Fig.\ref{fig_ratio}). Note that the spectral features of a fully grown pyramid are very similar to the spectral response from the hollow pyramid sides (see Supplementary Materials Fig.\ref{fig_ratio}). We conclude that both the peak intensity and the overall shape of the spectra are non-uniform along the WS$_2$ pyramid, behaving differently at the hollow pyramid crater and the stair-like sides. This indicates a difference in the atomic arrangement between the two parts of the pyramid.

\begin{figure*}[htp]
\centering
\includegraphics[width = 0.6\linewidth] {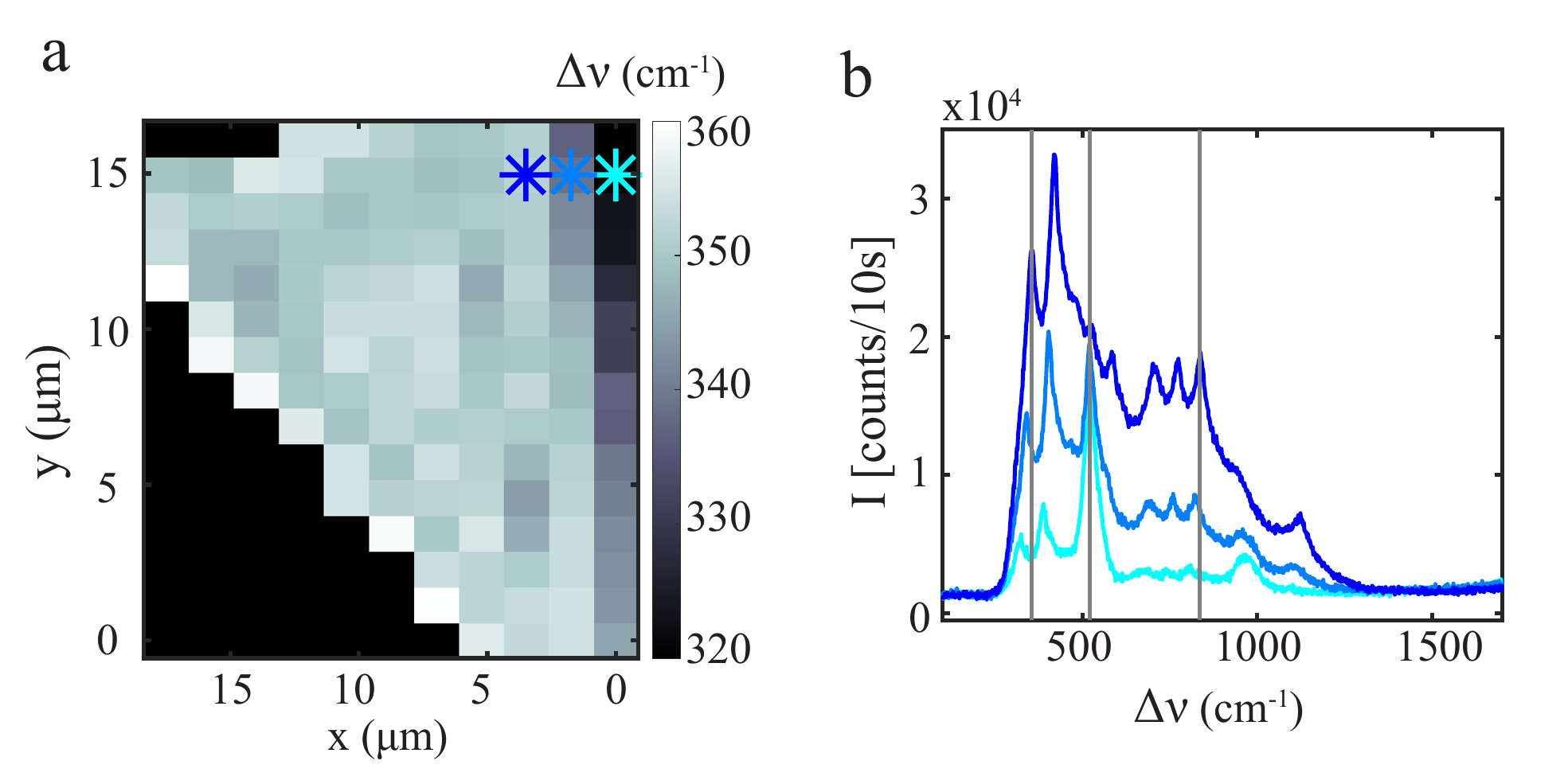}
\caption{\textbf{Position dependence of spectral position} \\ \textbf{a.} Map at \SI{4}{K} of the spectral position of the first WS$_2$ Raman feature (2LA(M),E$_{2g}$)\textcolor{black}{, with a step size of around \SI{1.5}{\mu m}}. The spectral position is fairly homogeneous over most of the pyramid, except for the edge, where it is blue shifted up to \SI{320}{cm^{-1}}. The stars indicate the position of the spectra in \textbf{b.} Three gray lines are drawn as guides to the eye. Comparing the position of the Si peak of the three spectra with the line at \SI{520}{cm^{-1}} indicates that the spectral position of this peak does not shift. Comparing the position of the first WS$_2$ peak with the line at \SI{355}{cm^{-1}} does indicate the large shift of the spectrum on the pyramid edge (light blue) with respect to the rest of the nanostructure.}
\label{fig_shift}
\end{figure*}

We also find a spatial non-uniformity in the spectral position of the peaks. \textbf{Figure \ref{fig_shift}a} depicts the spectral position of the first WS$_2$ Raman peak (E$_{2g}$,2LA) as a function of position. This spectral position is fairly uniform over most of the nanostructure, except for the right edge, where the spectral position is red shifted significantly up to \SI{320}{cm^{-1}}. Moreover, a small blue shift of the peak is seen on the left pyramid edge. The second WS$_2$ Raman peak exhibits a similar spectral shift (not shown), as do spectra obtained for a \SI{561}{nm} excitation. This spectral shift of peaks becomes more evident when comparing the spectra at the positions of the blue stars (Fig.\ref{fig_shift}b). The second line indicates the position of the silicon peak at \SI{520}{cm^{-1}}, which is clearly constant in all three spectra. However, when comparing the line at \SI{355}{cm^{-1}} with the position of the first Raman peak, it is clear that the WS$_2$ peaks in the spectrum are blue shifted. 

The spectral position of Raman modes in TMD materials is known to depend on the number of layers \cite{Lee_MoS2Ramanfirst_ACSNano_2010,  Zhao_RamanTMDlinear_Nanoscale_2013, Berkdemir_RamanWS2_ScientRep_2013, Zheng_MoS2pyramids_AdvMat_2017, Zhou_MoS2pyramid_Nanoscale_2018}. One might therefore have expected a gradual spectral shift along the stair-like sides of the pyramid, because of their gradual increase in WS$_2$ layer thickness. Unfortunately, the diffraction-limited laser spot of size \SI{450}{nm} (see Methods) is much bigger than the width of the individual terraces. Therefore, if the size of the steps is one or even a few layers, we do not have the resolution to distinguish thickness-dependent changes in the Raman response of individual steps. The changes in the Raman response are smallest for low N, where N is the number of layers. The reported difference in spectral position for different layer thicknesses is at most \SI{5}{cm^{-1}} (between a monolayer and a bilayer), much less than the shift of \SI{30}{cm^{-1}} that we observe at the edge of this pyramid. Therefore this spectral shift cannot be explained by a thickness increase alone. 

\begin{figure*}[htp]
\centering
\includegraphics[width = 0.6\linewidth] {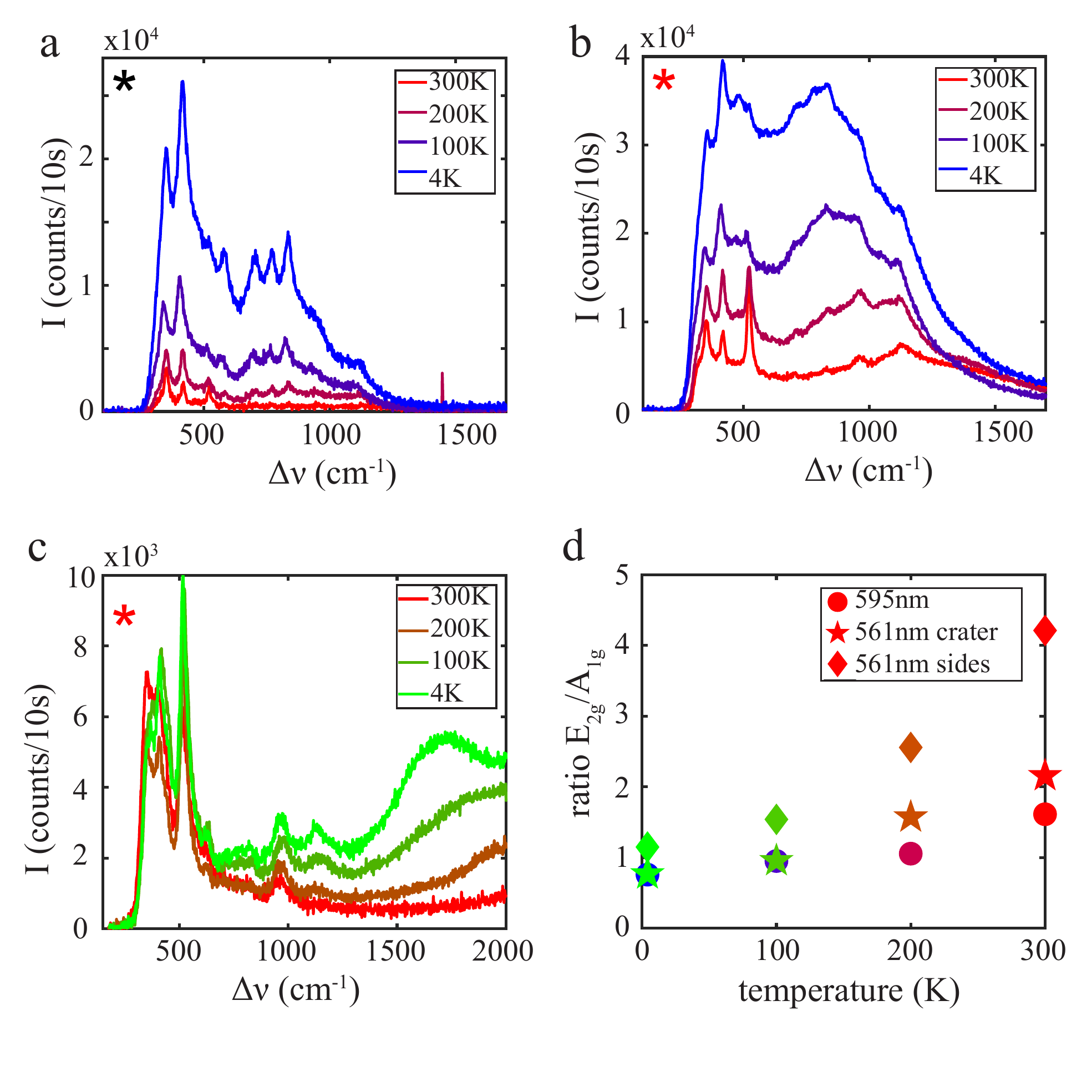}
\caption{\textbf{Temperature dependence of spectral features} \\ \textcolor{black}{\textbf{a-b.} Spectra obtained as a function of temperature with a \SI{595}{nm} excitation on the stair-like pyramid sides (black star in Fig.\ref{fig_position}) and on the hollow pyramid middle (red star in Fig.\ref{fig_position}), respectively. 
\textbf{c.} Spectra obtained with a \SI{561}{nm} excitation of the hollow pyramid middle as a function of temperature. The background visible at higher frequencies overlaps in wavelength with the background under the \SI{595}{nm} spectra in Fig.\ref{fig_position}b (see Supplementary Materials Fig.\ref{fig_background}) \textbf{d.} Temperature dependence of the intensity ratio of the first two Raman features, the fingerprint of WS$_2$ material. The intensity ratio is presented for spectra upon \SI{595}{nm} excitation (circles), spectra at the pyramid crater (stars) and pyramid sides (diamonds) upon \SI{561}{nm} excitation.}}
\label{fig_temperature}
\end{figure*}

The spectral position of Raman modes in TMD materials does not only depend on the number of layers, it is also known to be influenced by the defect density \cite{Mignuzzi_MoS2Raman_defectshift_PhysRevB_2015, Parkin_MoS2Raman_defectshift_ACSNano_2016}, strain \cite{Rice_MoS2Raman_strainshift_PRB_2013, Yang_MoS2Raman_strainshift_ScieRep_2014} and pressure \cite{Li_MoS2Raman_pressureshift_APL_2016}. The reported shift due to strain is 2-\SI{3}{cm^{-1}} \cite{Rice_MoS2Raman_strainshift_PRB_2013, Yang_MoS2Raman_strainshift_ScieRep_2014} and
due to an increased defect density is 5-\SI{10}{cm^{-1}} \cite{Mignuzzi_MoS2Raman_defectshift_PhysRevB_2015, Parkin_MoS2Raman_defectshift_ACSNano_2016}. For both strain and defects, the A$_{1g}$ peak is much less affected than the E$_{2g}$ peak. The reported shift due to pressure is up to \SI{40}{cm^{-1}} for pressures up to \SI{20}{GPa} \cite{Li_MoS2Raman_pressureshift_APL_2016}. 
Since our measurements were performed in either ambient (room temperature) or vacuum (low temperature) conditions, we do not expect a spectral shift due to pressure. It is not unlikely that a large defect density and/or the presence of strain are present in the hollow pyramids. Interestingly, spectra taken on fully grown WS$_2$ pyramids with curved edges do exhibit small shifts in the spectral peak position along the edges with highest curvature, where a higher stress or strain is expected (see Supplementary Materials Fig.\ref{fig_peak_pos}). Therefore we assume that the origin of the large spectral shift in the hollow WS$_2$ pyramid in Fig.\ref{fig_shift} lies in a combination of the mentioned effects of defect density and strain or stress. Having said that, the previously reported shifts, even when added, are much lower than the \SI{30}{cm^{-1}} that we observe on the edge of the hollow pyramid, so we cannot exclude unknown other causes related to the specific nanogeometry of the hollow pyramid. 

In this context, it is interesting to note that we measure an average spectral position of the first Raman peak on a WS$_2$ monolayer of \SI{357}{cm^{-1}}, which is higher than the average of \SI{350}{cm^{-1}} of this and other WS$_2$ pyramids (see Supplementary Materials Fig.\ref{fig_peak_pos}). Given that the first WS$_2$ Raman feature is a combination of the E$_{2g}$ and 2LA(M) phonon, we hypothesise that the first Raman feature in the monolayer has a larger contribution from the E$_{2g}$ than the same first feature in the hollow pyramid spectra. 

We conclude that the spectral features of the hollow pyramids, namely intensity, peak ratio and spectral peak position, vary in space over the nanostructures. This in contrast with the homogeneous distributions of these spectral features on a WS$_2$ monolayer. Moreover, the spectral position of the first WS$_2$ Raman feature is different for a WS$_2$ monolayer than for a hollow WS$_2$ pyramid, indicating a larger contribution from the E$_{2g}$ than the 2LA(M) phonon.

\subsection{Temperature dependence of spectral features}

Studying the temperature dependence of spectral features provides insights on the structural properties of the WS$_2$ pyramids. \textbf{Figure \ref{fig_temperature}a} presents spectra at four temperatures at the pyramid side (black star in Fig.\ref{fig_position}a), obtained with a \SI{595}{nm} excitation. With decreasing temperature, the Raman modes become more pronounced. Note for instance the three features at 702, 769 and \SI{833}{cm^{-1}}. The intensity of both the Raman features and the background increases with decreasing temperature. This intensity increase of the background is even more clear in Fig.\ref{fig_temperature}b-c, that present the spectra from the hollow pyramid middle (red star in Fig.\ref{fig_position}a) obtained upon either \SI{595}{nm} or \SI{561}{nm} excitation. The background also seems to exhibit a spectral blue shift as a function of temperature, with its maximum moving from \SI{635}{nm} at room temperature to \SI{620}{nm} at \SI{4}{K}. Based on the temperature dependence of its spectral position, we attribute this background to intermediate gap states or defect states rather than excitons, trions or an indirect bandgap response (see Supplementary Materials Section 3). 

\mbox{Figure \ref{fig_temperature}d} presents the temperature dependence of the average intensity ratio of the first two Raman peaks (E$_{2g}$/A$_{1g}$). As shown in Fig.\ref{fig_position}f, this ratio is not uniform over the pyramid, but is higher at the hollow pyramid middle than at the stair-like sides. This non-uniformity is most evident in the spectra obtained by \SI{561}{nm} excitation (diamonds in Fig.\ref{fig_temperature}d), as the A$_{1g}$ peak in the spectra from the pyramid sides almost completely disappears at room temperature (see Supplementary Materials Fig.\ref{fig_ratio}). For a \SI{595}{nm} excitation, at room temperature the first Raman peak ($E_{2g}$,2LA(M)) is 1.5x higher than the second Raman peak ($A_{1g}$) and at \SI{4}{K} this ratio is exactly inverted (circles in Fig.\ref{fig_temperature}d). The peak ratio upon \SI{561}{nm} excitation of spectra taken at the pyramid middle follow a similar temperature-dependent behaviour (stars in Fig.\ref{fig_temperature}d). 

The temperature-dependent intensity increase of TMDs Raman peaks has been reported previously for horizontal TMDs layers, and is attributed to an increase in phonon thermal population \cite{Fan_resonanceRamanTMD_JApplPhys_2014, Gaur_temperatureRamanWS2_PhysChemC_2015, Peimyoo_temperatureRamanWS2_NanoRes_2015}. The difference in intensity ratio between the E$_{2g}$ and A$_{1g}$ for the different excitation frequencies can be explained by the more resonant \SI{595}{nm} laser exciting the Raman peaks differently than the out-of-resonance \SI{561}{nm} laser \cite{Carvalho2015, Corro_resonantRamanTMD_NanoLetters_2016}. \textcolor{black}{The strength of the exciton-phonon interaction, and therefore the resonance condition, is different for the in-plane E$_{2g}$ than the out-of-plane A$_{1g}$ Raman modes \cite{Mastrippolito_excitonPhonon_Nanoscale_2020, Corro_resonantRamanTMD_NanoLetters_2016}. This explains why the ratio E$_{2g}$/A$_{1g}$ is higher for a \SI{561}{nm} excitation, \textit{e.g.}, out-of-resonance with the excitonic transition, than for the resonant \SI{595}{nm} excitation. The Raman intensity ratio also depends on the layer thickness of the material \cite{Zhao_RamanTMDlinear_Nanoscale_2013}. However, the main difference in intensity ratio is reported between a monolayer and a bilayer, whereas we observe different relative Raman intensities between the few-layer pyramid crater and the thick pyramid edge.} Moreover, the temperature-dependent behaviour of the Raman intensity also depends on the defect density in the sample \cite{Li_RamanWS2_defectsT_JPhysChemC_2019}. These factors are not mutually independent, \textit{e.g.}, \textcolor{black}{the WS$_2$ thickness influences the exciton-phonon interaction,} the resonance of the excitation affects the influence of phonons and defects on the Raman intensity. The temperature dependence of the Raman peak intensities, excited with two different frequencies, is therefore an interplay between the phonon thermal population, the resonance conditions for the different phonon peaks and the defect density in the structure. 

It is interesting to note in this context the intensity ratio of the first two Raman peaks on a WS$_2$ monolayer, excited at \SI{561}{nm}. This ratio is 1.4 at room temperature, much lower than both the intensity ratio at the hollow pyramid middle and at the stair-like sides, excited at \SI{561}{nm} (see Supplementary Materials Fig.\ref{fig_ratio}). Unexpectedly, this indicates a clear difference in structure and/or thickness between the WS$_2$ monolayer and hollow pyramid. The large difference in intensity ratio on the hollow pyramid middle and the stair-like sides suggests a difference in atomic arrangement, pointing out that nanogeometrical differences induce spectral modifications.

\section{Conclusions}

We have studied the optical response of hollow WS$_2$ pyramids, comparing them with WS$_2$ monolayers grown on the same substrate. The optical response of these nanostructures is completely different, as hollow WS$_2$ exhibit a strongly reduced photoluminescence with respect to WS$_2$ monolayers. This enables us to study the rich variety of Raman peaks that the pyramids exhibit as a result of the resonant excitation. Following the hypothesis of a multiphonon excitation involving the longitudinal acoustic phonon LA(M), we are able to explain the origin of all 10-12 observed Raman resonances. 
In contrast with monolayers, the measured optical response of the pyramids is non-uniform in both intensity, intensity ratio between peaks, spectral shape and spectral position. We attribute the spectral differences between the hollow pyramid middle and the stair-like sides to differences in both nanogeometry and atomic arrangement. ADF-STEM measurements confirm variations in the atomic arrangement, where the level of disorder is more marked in the pyramid crater than on the sides. Next to a positional dependence, we measure the temperature dependent behaviour of the spectral response of the hollow WS$_2$ pyramids. With decreasing temperature, the spectra change in intensity and shape. We see clear differences between spectra obtained with a resonant and out-of-resonance excitation laser. As the optical response of WS$_2$ monolayers, exhibiting photoluminescence, is completely different, we therefore deduce to have fabricated a platform of structures with tunable optical properties. Both nanostructures offer exciting possibilities, with applications ranging from opto-electronics to non-linear optics.  

\section{Methods} \label{methods}

The WS$_2$ hollow pyramids are directly grown on a microchip using chemical vapour deposition (CVD) techniques. This microchip is composed of a silicon frame with nine windows over which a continuous silicon nitride (Si$_3$N$_4$) film is spanned. Preceding the CVD growth procedure\cite{Sabrya2020}, tungsten trioxide (WO$_3$) is deposited onto the microchip. This deposition is achieved by dispersing \SI{50}{mg} of WO$_3$ in \SI{1}{mL} of isopropanol, followed by a deposition by means of a pipette. The subsequent sulfurization process \cite{Sabrya2020} is carried out in a gradient tube furnace from Carbolite Gero. The microchip is placed in the middle zone and a crucible holding \SI{400}{mg} of sulfur is placed upstream from it. The middle zone is heated to 750\textcelsius{}, and kept at this reaction temperature for 1 hour after which the system is naturally cooled down to room temperature. Sulfurization is carried out under an argon flow of \SI{150}{sccm}. The zone containing the sulfur is heated to 220\textcelsius{}.

The optical measurements are performed using a home-built spectroscopy set-up. We placed the sample in a Montana cryostation S100, using Attocube ANPxy101/RES piezo scanners to perform the presented raster scans. The small cross-coupling between the in-plane and out-of-plane piezo scanners causes a skewing between the x and y axis of the spectral-feature maps depicted in this work. Comparing the maps with the optical and SEM images of the pyramids, we can correlate the position of certain spectral features with the position on the pyramid. The cryostation is cooled down from room temperature to \SI{200}{K}, \SI{100}{K} and \SI{4}{K}. The sample is illuminated through an \SI{0.85}{NA} Zeiss 100x objective. Two continuous wave lasers are used, one with a wavelength of \SI{595}{nm} and a power of \SI{1.6}{mW/mm^2}, and one with a wavelength of \SI{561}{nm} and a power of \SI{3.6}{mW/mm^2}. To avoid the consequences of tight focussing on (circular) polarization, a \SI{2}{mm} laser diameter is used, slightly underfilling the objective in the excitation path. \textcolor{black}{This results in an excitation spot size of approximately \SI{450}{nm}}. A linear (vertical) polarization is used for the laser light. The sample emission is collected in reflection through the same objective as in excitation, and projected onto a CCD camera (Princeton Instruments ProEM 1024BX3) and spectrometer (Princeton Instruments SP2358) \textit{via} a 4f lens system. The excitation light is filtered out using colour filters. 

The transmission electron microscopy (TEM) measurements were carried out using an ARM200F Mono-JEOL microscope with Cs probed corrected. The microscope was operated at \SI{200}{kV} both in TEM and STEM modes, with the monochromator on and a slit of \SI{2}{\mu m} inserted. For the atomic resolution ADF-STEM measurements, an objective aperture of \SI{30}{\mu m} and a camera length of \SI{12}{cm} were used. The convergence semi-angle was \SI{23}{mrad}.

\section{Acknowledgements} 
The authors acknowledge funding from ERC Starting Grant “TESLA” No. 805021. The authors acknowledge dr Martin Caldarola and dr Filippo Alpeggiani for their help in the data analysis. 

\clearpage

\section*{Supplementary Materials}

\subsection{WS$_2$ hollow pyramid sample}

\begin{figure*}[htp]
\centering
\includegraphics[width = 0.6\linewidth] {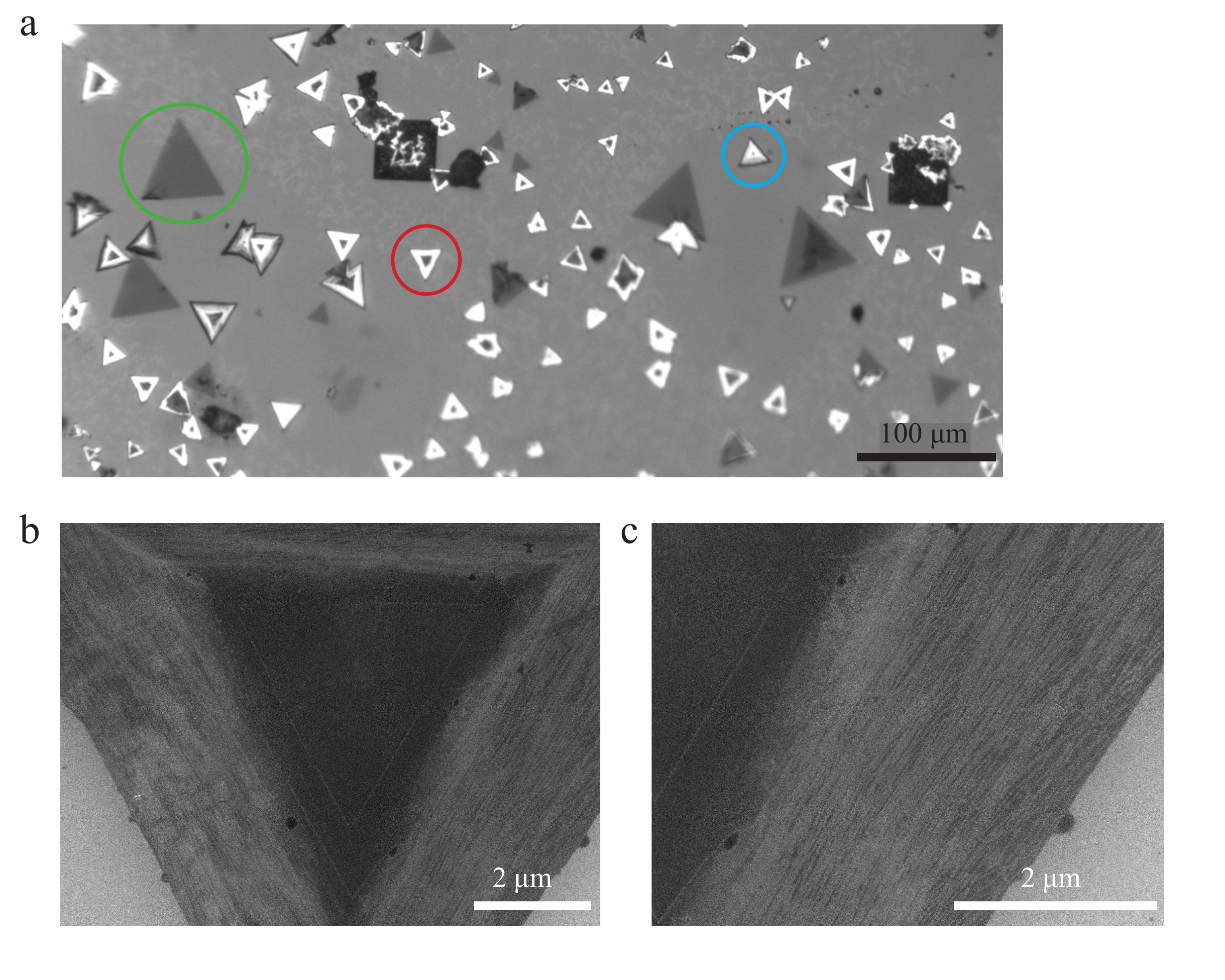}
\caption{\textbf{Hollow WS$_2$ pyramid sample} \\ \textbf{a.} Wide-field optical image of a part of the sample, with hollow pyramids (red), WS$_2$ monolayer flakes (green) and full pyramids (blue). The black squares are windows in the silicon frame over which a silicon nitride film is spanned (see Methods). Under these growth conditions, many hollow pyramids arise size 10 - \SI{25}{\mu m}, comparable to the one presented in this work. \textbf{b,c.} SEM images of the hollow WS$_2$ pyramid studied in the main text. In \textbf{b}, the black triangle in the middle is the bottom of the pyramid crater. The top rim of the pyramid can also be distinguished around the triangle of the pyramid crater. The steps in the stair-like sides can be easily recognized in \textbf{c}. }
\label{fig_microscope}
\end{figure*}

Figure \ref{fig_microscope}a presents a wide-field optical image of a part of the sample. Under these growth conditions both hollow pyramids (red), monolayer flakes (green) and full pyramids (blue) are created. Note that there are many hollow pyramids with a comparable size to the one studied in this work (\SI{15}{\mu m}). Using the same CVD growing conditions (see Methods) yields similar samples with the same distribution of pyramid-like structures.

To provide better insight in the morphology of the hollow WS$_2$ pyramids, Fig.\ref{fig_microscope}b,c depict higher magnification SEM images of the hollow pyramid depicted in Fig.1a in the main text. The black triangle in the middle of Fig.\ref{fig_microscope}b is the bottom of the pyramid crater. The top rim can also be distinguished around the crater triangle. The steps of the stair-like sides can clearly be recognized in Fig.\ref{fig_microscope}c.

\begin{figure*}[hbp]
\centering
\includegraphics[width = 0.7\linewidth] {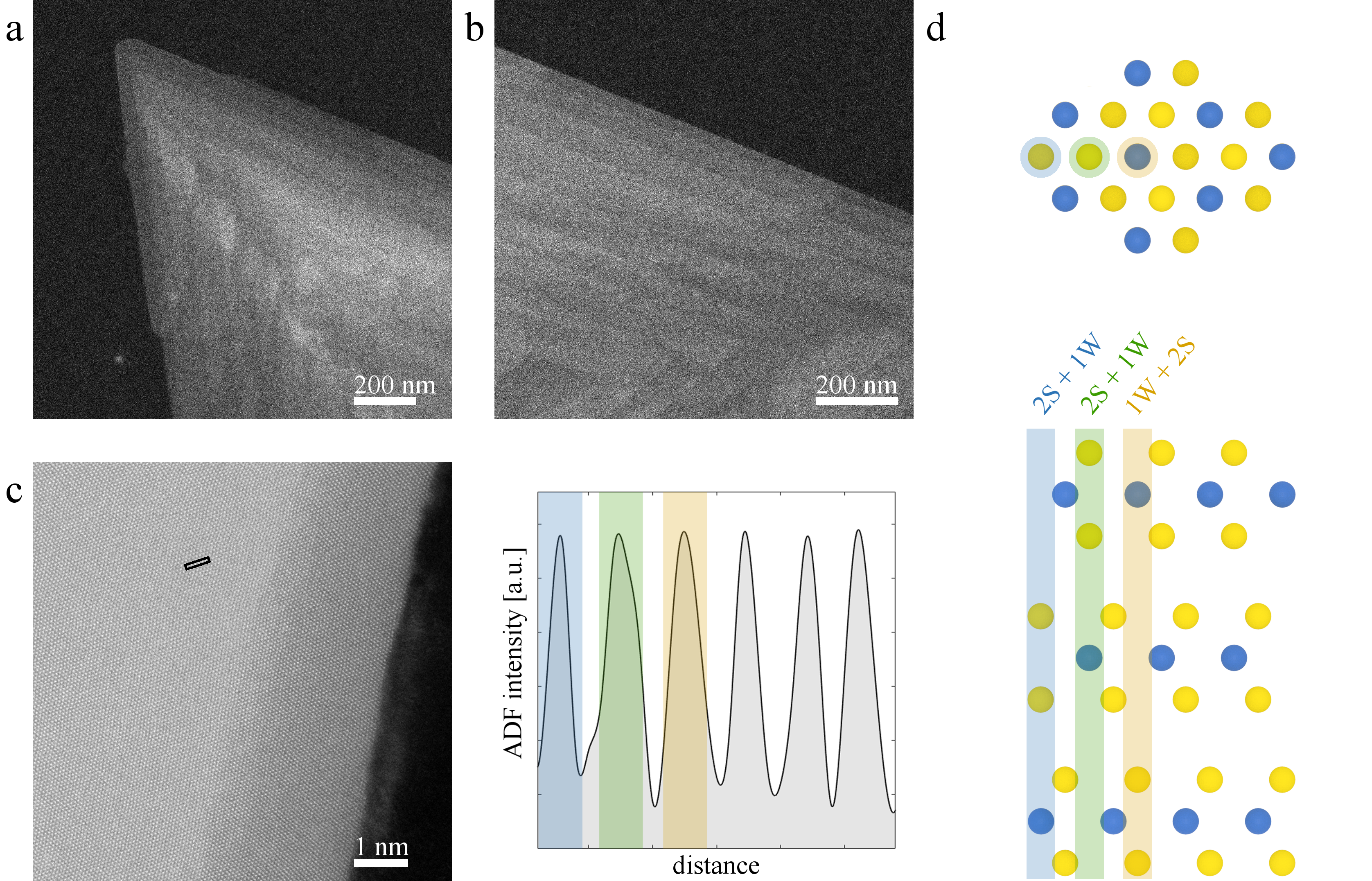}
\caption{\textbf{Morphology and crystal structure of the hollow WS$_2$ pyramids} \\ \textbf{a-b.} Low-magnification ADF-STEM images of a hollow WS$_2$ pyramid. The step-like nature of the pyramid side is clearly visible by the changes in contrast with every step. \textbf{c.} Atomic resolution image corresponding to the side of the hollow pyramid (left panel) and the ADF intensity profile (right panel) acquired along the black outlines region in the atomic resolution image. \textbf{d.} Schematic atomic model of the top-view (upper panel) and side-view (lower) of the crystalline structure associated to the 3R-WS$_2$ phase.}
\label{fig_TEM-morp-and-3R}
\end{figure*}

In addition, Transmission Electron Microscopy (TEM) measurements are performed to gain access to the atomic structure of the hollow pyramids. Figure \ref{fig_TEM-morp-and-3R}a and \ref{fig_TEM-morp-and-3R}b display low-magnification annular dark-field (ADF) scanning transmission electron microscopy (STEM) images of the side of a hollow WS$_2$ pyramid. The variations in the contrast visualise clearly the step-like nature of the hollow pyramid side. Figure \ref{fig_TEM-morp-and-3R}c presents an atomic-resolution ADF-STEM image corresponding to the side of the pyramid. Each bright spot corresponds to an atomic column that is composed of alternating tungsten (W) and sulfur (S) atoms. Using an ADF linescan, extracted from the atomic resolution image across six lattice points, we confirm that the WS$_2$ within the hollow pyramid crystallizes in a 3R crystal phase (see Fig.\ref{fig_TEM-morp-and-3R}d).

\subsection{Spectral background}

\begin{figure*}[htp]
\centering
\includegraphics[width = \linewidth] {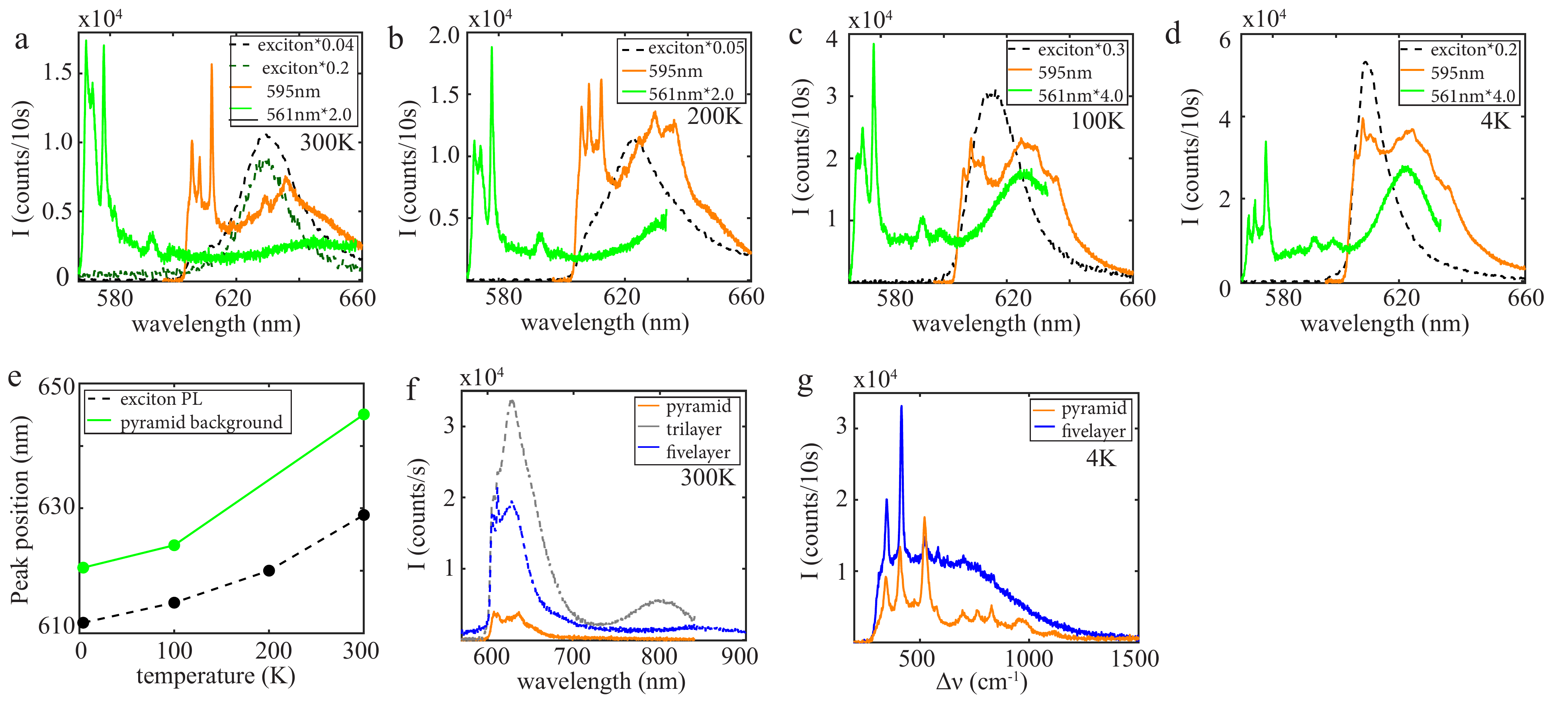}
\caption{\textbf{Comparison pyramid, monolayer and few-layer spectra} \\ \textcolor{black}{\textbf{a-d.} Spectral response of a pyramid upon a \SI{595}{nm} excitation (orange) and a \SI{561}{nm} excitation (green), and the spectral response of a monolayer upon \SI{595}{nm} excitation (black dotted line), at temperatures between room temperature and \SI{4}{K} (the spectra are re-scaled for easier comparison, see legends). The spectral response of the monolayer upon a \SI{561}{nm} excitation (green dotted line in \textbf{a}) has a lower intensity than to the \SI{595}{nm} excitation, but is at the same wavelength. \textbf{e.} The background under the higher order Raman features in the spectra of the \SI{595}{nm} excitation, and the spectral position of the exciton PL are both blue-shifting with decreasing temperature. \textbf{f.} In contrast to the spectra of a WS$_2$ trilayer (in grey) and five layers of WS$_2$ (in blue) (exfoliated on a Si substrate), the spectrum of the WS$_2$ pyramid (in orange) does not exhibit light from an indirect bandgap at 800 - \SI{850}{nm} wavelength. Moreover, although reduced with respect to the monolayer, the PL from the direct bandgap of the few-layers of WS$_2$ is clearly distinguishable from the background. \textbf{g.} Comparison of the spectral response of the hollow pyramid (in orange) and of five layers of WS$_2$ (in blue), acquired at \SI{4}{K} with a \SI{595}{excitation}. }}
\label{fig_background}
\end{figure*}

The hollow pyramid spectra exhibit a number of Raman modes plus a background (see Fig.3, Fig.6 in the main text). Using the temperature-dependent spectral position of this background, we can explain its origin. \mbox{Figure \ref{fig_background}} depicts the spectral response of the hollow pyramid upon a \SI{595}{nm} excitation (orange) and a \SI{561}{nm} excitation (green), comparing this with the photoluminescence of a monolayer exciton (black dotted line). The spectral response of the hollow pyramid to a \SI{595}{nm} excitation exhibits a background under the higher order Raman modes, whereas the spectral response of a \SI{561}{nm} excitation exhibits a background separated from the Raman modes, that overlap spectrally (see Fig.\ref{fig_background}\textcolor{black}{c-d}). The spectral position of the photoluminescence peak is completely different (\mbox{610 - \SI{630}{nm}} in the temperature range \SI{4}{K} - \SI{300}{K}). Figure \ref{fig_background}\textcolor{black}{e} depicts the spectral position of the pyramid background, determined from the spectra of \SI{561}{nm} excitation (in green), and the spectral position of the exciton PL (black dotted line). Both the pyramid background and the exciton PL peak are blue-shifting with decreasing temperature, however their spectral position is different from each other. At room temperature (Fig.\ref{fig_background}a), the spectral position of the PL peak (at \SI{630}{nm}) is very close to the background of the pyramid spectra acquired with a \SI{595}{nm} excitation (around \SI{645}{nm}). At \SI{4}{K} (Fig.\ref{fig_background}\textcolor{black}{d}), the PL peak (at \SI{615}{nm}) overlaps roughly with the first few Raman features of the pyramid spectra acquired with a \SI{595}{nm} excitation (around \SI{620}{nm}). However, at \SI{200}{K} and especially \SI{100}{K} (Fig.\ref{fig_background}\textcolor{black}{b,c}), the PL peak overlaps neither with the first Raman features nor the background of the pyramid spectra acquired with a \SI{595}{nm} excitation. Therefore we conclude that the background of the hollow pyramid spectra is not photoluminescence from the direct bandgap of WS$_2$. 

Another potential explanation for the background in the pyramid spectra is emission from the indirect bandgap. Few-layer WS$_2$ samples exhibit a combination of direct and indirect bandgap emission \cite{Mak_MoS2monofirst_PhysRevLett_2010}. Figure \ref{fig_background}f compares the room-temperature spectra of a WS$_2$ trilayer (grey) and five WS$_2$ layers (blue) (exfoliated on a Si substrate) with a spectrum from the hollow WS$_2$ pyramid. The room-temperature spectral position of the indirect bandgap ranges from \SI{700}{nm} for a bilayer to \SI{850}{nm} for multilayers \cite{SuHyun_2018}. This constitutes a large spectral separation with the measured spectrum (compare Fig.\ref{fig_background}a). Moreover, with decreasing temperature, the indirect bandgap is reported to exhibit a red shift, \textit{i.e.}, away from the exciton position \cite{Molas_PLindirectTemp_Nanoscale_2017, Zhao_PLIndirectTemp_NanoLett_2013}, whereas the pyramid background exhibits a similar blue shift to the exciton PL (see Fig.\ref{fig_background}\textcolor{black}{e}).   

An alternative explanation would be emission from a charged exciton or trion. However, the reported spectral position of the trion lies closer to the exciton than the background in the hollow pyramid spectra  \cite{Plechinger_WS2trionsTemp_PSS_2015,  Kato_WS2trions_ACSNano_2016}. We conclude that this background originates in intermediate gap states or defect states, that are reported to be in a range of spectral positions further away from the exciton PL than the trion \cite{Jadczak_TMDtrionsTemp_Nanotech_2017, Kato_WS2trions_ACSNano_2016, He_WS2defect_ACSNano_2016}, but closer than the indirect bandgap \cite{SuHyun_2018, Molas_PLindirectTemp_Nanoscale_2017, Zhao_PLIndirectTemp_NanoLett_2013}.

It is interesting to note that the spectral background from these intermediate gap states is more present in the pyramid crater than on the pyramid sides. We propose this might be originated in the higher density of crystallographic defects. 

\begin{figure*}[htp]
\centering
\includegraphics[width = 0.9\linewidth] {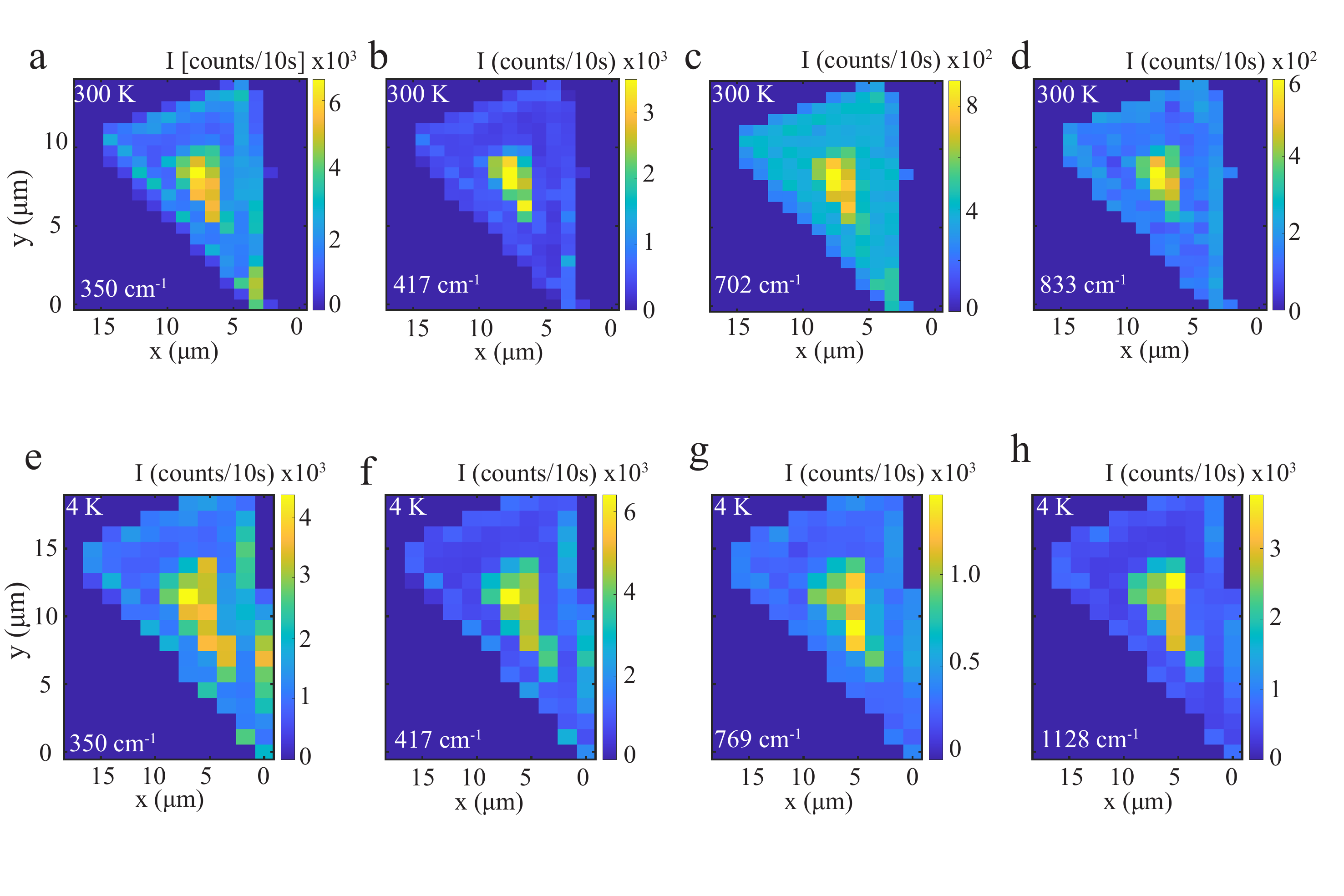}
\caption{\textbf{Position dependence of Raman intensity} \\ \textbf{a-d} Intensity map of the Raman features of the hollow WS$_2$ pyramid \textbf{a.} around \SI{350}{cm^{-1}} (2LA,E$_{2g}$), \textbf{b.} around \SI{417}{cm^{-1}} (A$_{1g}$), \textbf{c.} around \SI{702}{cm^{-1}} (4LA) and \textbf{d.} around \SI{833}{cm^{-1}} (2A$_{1g}$), taken at room temperature upon a \SI{561}{nm} excitation. \textbf{e-h} Intensity map of the Raman features of the hollow WS$_2$ pyramid \textbf{e.} around \SI{350}{cm^{-1}} (2LA,E$_{2g}$), \textbf{f.} around \SI{417}{cm^{-1}} (A$_{1g}$), \textbf{g.} around \SI{770}{cm^{-1}} (A$_{1g}$+2LA) and \textbf{h.} around \SI{1120}{cm^{-1}} (6LA), taken at \SI{4}{K} upon a \SI{561}{nm} excitation. Note that in all cases (compare with Fig.4a,b in the main text) the intensity of the Raman features from the pyramid crater is significantly higher than the Raman intensity from the stair-like sides.}
\label{fig_intensity}
\end{figure*}

\begin{figure*}[htp]
\centering
\includegraphics[width = 0.9\linewidth] {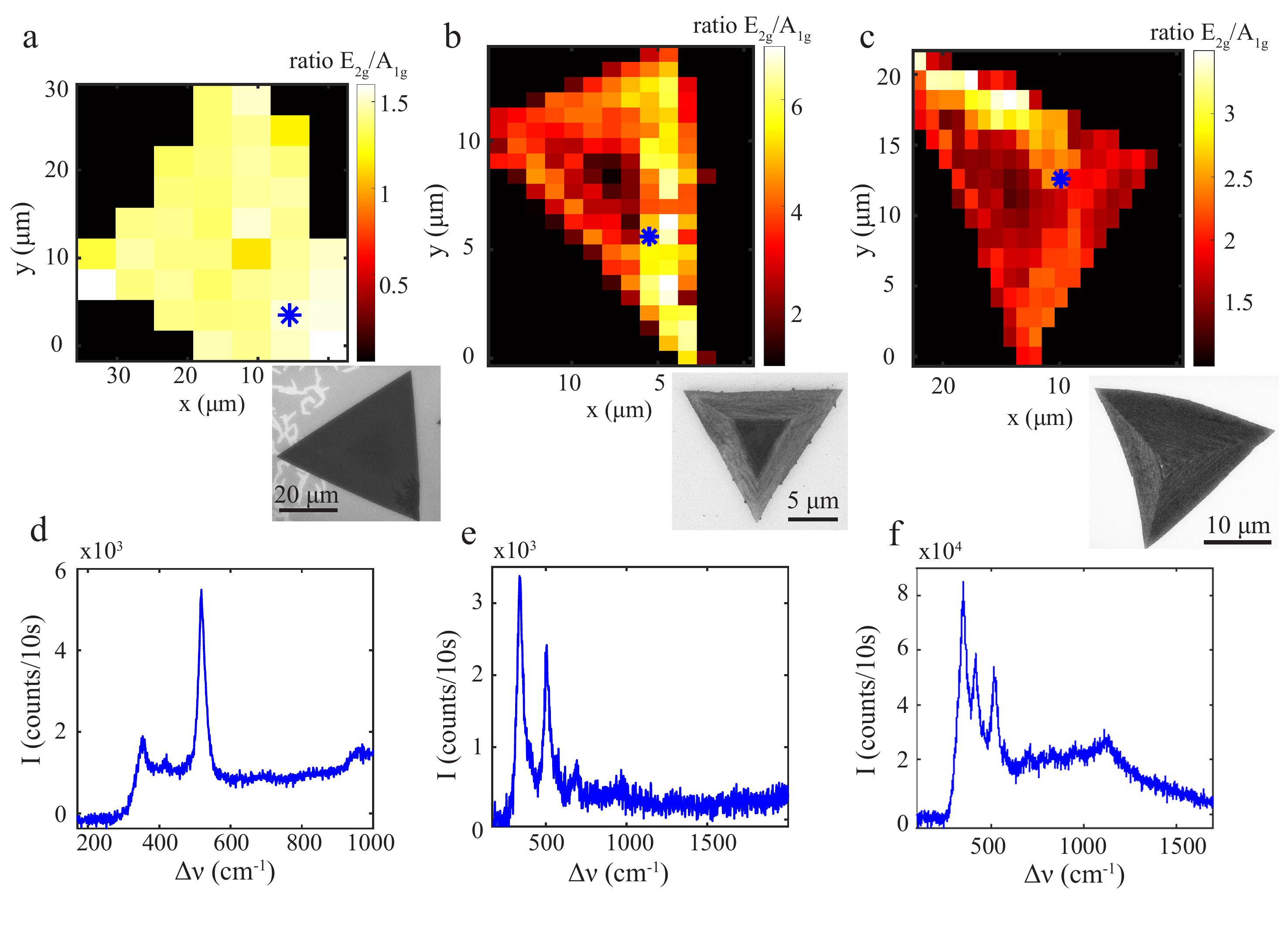}
\caption{\textbf{Position dependence of peak ratio} \\ \textbf{a-c} Room-temperature map of the ratio between the first two Raman features in the spectra (E$_{2g}$/A$_{1g}$) of \textbf{a} the monolayer, \textbf{b} the hollow pyramid and \textbf{c} a full WS$_2$ pyramid (compare SEM images) (blue stars indicate the positions of the spectra in \textbf{d-f}). \textbf{a.} The peak ratio on the monolayer is homogeneous along the full flake, 1.4 on average (upon a \SI{561}{nm} excitation). \textbf{b.} The peak ratio on the hollow pyramid (upon a \SI{561}{nm} excitation) is much higher along the stair-like pyramid sides (namely 4-7) than on the pyramid crater (roughly 1.0). \textbf{c.} The peak ratio on the full pyramid is significantly higher along the upper side. Note in the presented SEM image that this side of this full WS$_2$ pyramid is curved, potentially inducing strain. \textbf{d.} Raman spectrum of the WS$_2$ monolayer. \textbf{e.} Spectrum of the side of the WS$_2$ hollow pyramid upon \SI{561}{nm} excitation. Note that the A$_{1g}$ feature is merely a shoulder on the E$_{2g}$,2LA(M) feature, which explains the high peak ratio in \textbf{b}. \textbf{f.} Spectrum of the full WS$_2$ pyramid (\SI{595}{nm} excitation), which is comparable to the spectra of the hollow WS$_2$ pyramid presented in the main text.}
\label{fig_ratio}
\end{figure*}

\textcolor{black}{\subsection{Spectral features of pyramids, monolayer and few-layer WS$_2$}}


Figure \ref{fig_intensity} presents maps of the intensity of different Raman features in spectra from the hollow WS$_2$ pyramid, taken at room temperature (Fig.\ref{fig_intensity}a-d) and at \SI{4}{K} (Fig.\ref{fig_intensity}e-h) upon a \SI{561}{nm} excitation. Note that for all the Raman features the intensity from the pyramid crater is higher than from the pyramid sides, as was the case for the Raman features depicted in Fig.4a,b in the main text (\SI{595}{nm} excitation, \SI{200}{K}). We conclude that the intensity distribution of the Raman features is independent of temperature or excitation frequency. As alluded to in the main text, we hypothesise that light scatters from the stair-like pyramid sides and thus reduces the available excitation light to excite any Raman modes, or that scattering of the resulting Raman response reduces the amount of light detected. 

\textcolor{black}{Figure \ref{fig_background}g depicts a comparison between the spectral response of the pyramid (in orange) and five layers of exfoliated WS$_2$ (in blue) at \SI{4}{K} and upon a \SI{595}{nm} excitation. The spectral response of the few-layer WS$_2$ exhibits a combination of Raman modes and PL from the excitonic resonance. Due to the high PL intensity, the higher-order Raman modes are much less clear in the few-layer WS$_2$ than in the pyramid spectrum. In both spectra, the A$_{1g}$ mode has a higher intensity than the E$_{2g}$ mode. The ratio of E$_{2g}$/A$_{1g}$ is 0.60 for few-layer WS$_2$, whereas it is on average 0.79 for the pyramid spectra (see Fig.6b in the main text). }

Figure \ref{fig_ratio}a-c present room-temperature maps of the ratio between the first two Raman peaks in the spectra (E$_{2g}$/A$_{1g}$) of respectively the WS$_2$ monolayer, the hollow pyramid and a full WS$_2$ pyramid (compare the presented SEM images). In contrast with the position dependent peak ratio of the hollow pyramids, the peak ratio on the monolayer is homogeneous along the full flake (see Fig.\ref{fig_ratio}a). This ratio is around 1.4, as can be observed in the Raman spectrum in Fig.\ref{fig_ratio}d. The Raman peak ratio of the hollow pyramids upon \SI{561}{nm} excitation is much higher on the pyramid sides than in the pyramid crater (see Fig.\ref{fig_ratio}b). This difference in ratio was already apparent upon \SI{595}{nm} excitation, as presented in Fig.4f in the main text, but the contrast between the pyramid sides and crater is much larger upon \SI{561}{nm} excitation. The Raman peak ratio at the pyramid sides is as high as 4-7 times, as the A$_{1g}$ feature is reduced to merely a shoulder on the E$_{2g}$,2LA(M) feature. This becomes apparent in the spectrum in Fig.\ref{fig_ratio}e. The SEM image under Fig.\ref{fig_ratio}c depicts a full WS$_2$ pyramid grown on the same substrate. This pyramid exhibits a clear curvature. Interestingly, the Raman peak ratio of the spectra from this pyramid (upon a \SI{595}{nm} excitation) exhibit a difference on the side with the largest curvature, potentially induced by strain. Figure \ref{fig_ratio}f presents a spectrum of the full WS$_2$ pyramid, which is similar to the spectra of the hollow WS$_2$ pyramid in the main text. We conclude that the Raman peak ratio provides information on differences in the atomic structure between different nanostructures, as well information on differences in both atomic structure and strain or stress within the same nanostructure. 

\begin{figure*}[htp]
\centering
\includegraphics[width = 0.9\linewidth] {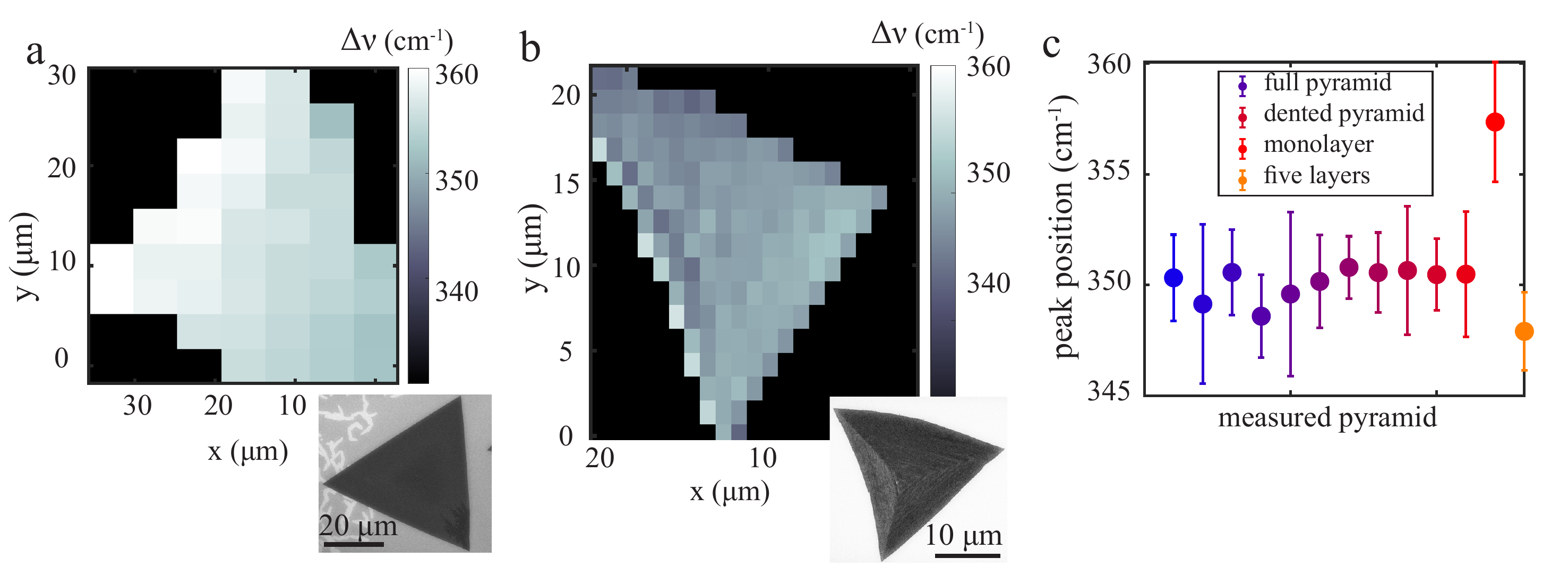}
\caption{\textbf{Position dependence of spectral position} \\ \textbf{a-b} Room temperature map of the first WS$_2$ Raman feature around 350cm$^{-1}$ on \textbf{a} the monolayer and \textbf{b} a full WS$_2$ pyramid (compare SEM images). Comparing \textbf{a} and \textbf{b} and Fig.5 in the main text, it becomes apparent that the Raman peak position on the monolayer is significantly different than on the pyramids. \textbf{c} depicts the average spectral position of the first Raman peak on all measured pyramids and the monolayer. Note that the bar does not indicate errors, but rather the spread of the peak position along the pyramids. The first Raman feature on the monolayer is around \SI{357}{cm^{-1}}, whereas on all measured pyramids, it is around \SI{350}{cm^{-1}}\textcolor{black}{, and around \SI{348}{cm^{-1}} on five layers of WS$_2$}. Possibly this Raman peak has a larger contribution from the E$_{2g}$ than the 2LA(M) phonon in the monolayer than in the pyramid spectra.}
\label{fig_peak_pos}
\end{figure*}

Another position-dependent spectral feature alluded to in the main text is the spectral position of the Raman peaks. Figure \ref{fig_peak_pos} presents a map of the spectral position of the first Raman peak on the WS$_2$ monolayer and the full WS$_2$ pyramid respectively. The peak position is fairly homogeneous along both nanostructures compared to the hollow pyramid (Fig.5a in the main text). Figure \ref{fig_peak_pos}c depicts the average spectral position of the first Raman peak, as measured on different WS$_2$ pyramids and the monolayer (the bar does not indicate errors, but rather the spread of the peak position along the pyramids). The peak position on the monolayer is around \SI{357}{cm^{-1}}, which is significantly different than the peak position of around \SI{350}{cm^{-1}} on all the pyramid structures \textcolor{black}{and \SI{348}{cm^{-1}} for five layers of exfoliated WS$_2$}. We hypothesise that this first Raman peak has a larger contribution from the E$_{2g}$ than the 2LA(M) phonon in the monolayer than in the pyramid spectra. In conclusion, the different nanogeometry of the WS$_2$ pyramids induces spectral changes with respect to monolayer WS$_2$.\\

\subsection{Structural characterisation}

\begin{figure*}[htp]
\centering
\includegraphics[width = 0.8\linewidth] {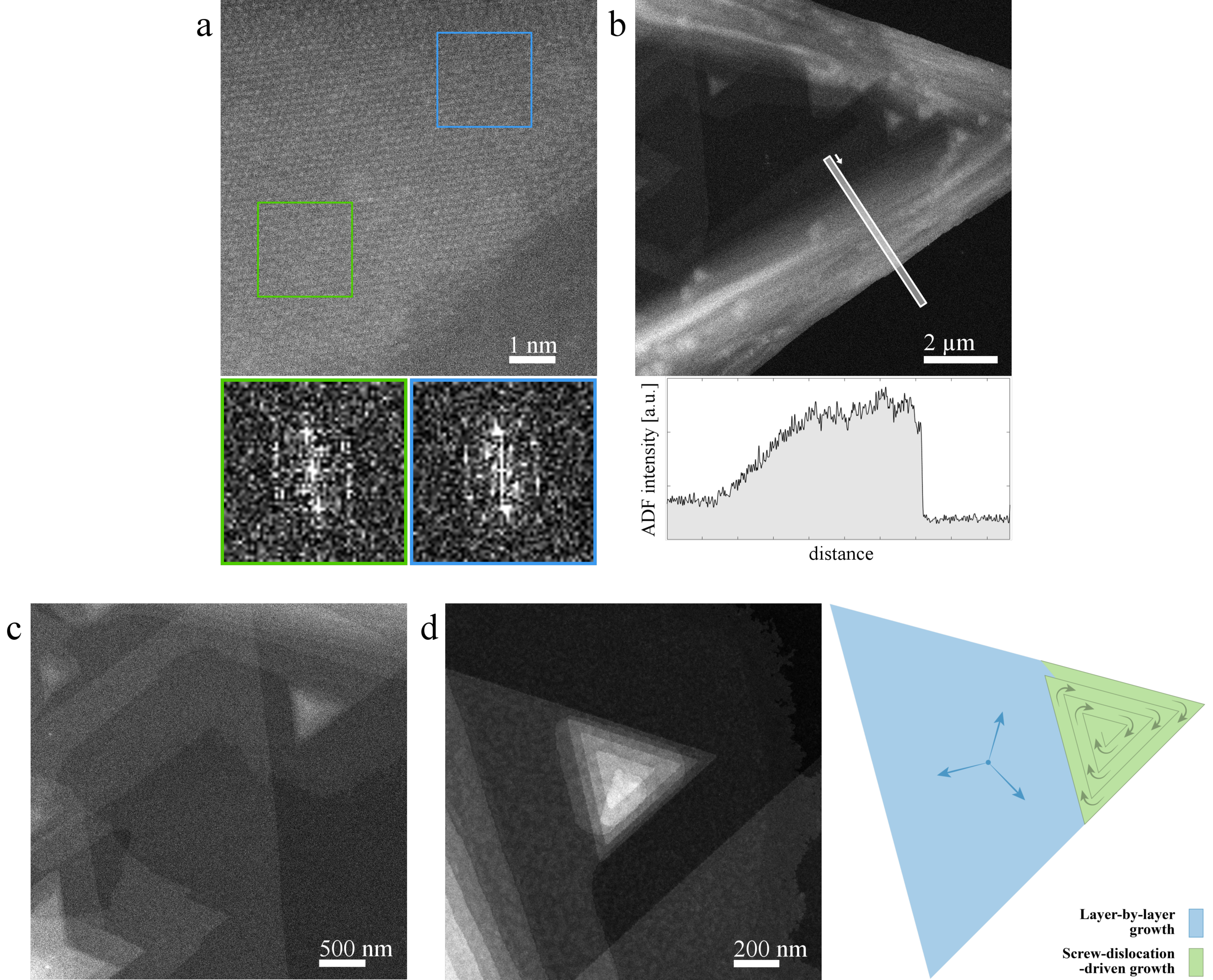}
\caption{\textbf{ADF-STEM of a hollow WS$_2$ pyramid} \\ \textbf{a.} Atomic resolution ADF-STEM image taken at the side of a hollow WS$_2$ pyramid. \textbf{inset} FFTs are taken within two visual different areas of this given region, highlighting a clear difference between the two areas. \textcolor{black}{\textbf{b-d.}} Low-magnification ADF images of the same hollow pyramid. The characteristic morphology of the hollow pyramids can clearly be observed, and is also extracted from the line profile over the white line given within \textbf{b} (see inset). \textcolor{black}{Each of these images show free-standing flakes which arise from the walls of the hollow pyramid. The free-standing flake shown in \textbf{d} suggest a growth mechanism where layer-by-layer stacking and screw-dislocation-driven growth mechanisms co-exist. \textbf{inset} A schematic visualises this co-existence. Here a free-standing flake initiates its growing from a central nucleation point via layer-by-layer growth (blue), after which a separate nucleation event occurs leading to a screw-dislocation-driven growth (green) on top.}}
\label{fig_additional_TEM}
\end{figure*}

Figure \ref{fig_additional_TEM} displays additional results obtained in the structural characterization measurements by means of Transmission Electron Microscopy (TEM). Figure \ref{fig_additional_TEM}a presents a high-resolution ADF-STEM image taken at the side of a hollow WS$_2$ pyramid. Within Figure 2 of the main text it could already be observed that the atomic arrangements present in the sides of the WS$_2$ pyramid and the middle are not the same. Figure \ref{fig_additional_TEM}a depicts how even within a given region the atomic arrangement can vary. This is highlighted by the differences in the FFTs taken in the lower left corner (marked by the green rectangle in Fig.\ref{fig_additional_TEM}a) and the top right corner (marked by the blue rectangle in Fig.\ref{fig_additional_TEM}b) of this given region. Where the FFT in the blue area shows two nicely arranged hexagonal patterns, \textit{e.g.}, an inner and an outer hexagon, the green area presents one hexagon. These subtle variations of the atomic arrangement might be induced by the local presence of strain, which in turn results into a slight change of the orientation of the flake.

Figures \ref{fig_additional_TEM}b \textcolor{black}{to \ref{fig_additional_TEM}d} depict low-magnification ADF images of the hollow WS$_2$ pyramid. Within Fig.\ref{fig_additional_TEM}b the hollow nature of the pyramid can clearly be observed, and illustrated even clearer in the line profile taken along the white line given in Fig.\ref{fig_additional_TEM}b. In addition, \textcolor{black}{Fig.\ref{fig_additional_TEM}b-d} depict how free-standing WS$_2$ flakes seem to arise from the walls of the hollow pyramid. These increase the level of structural disorder in the middle of the pyramids as compared to the sides. \textcolor{black}{The standing flake depicted in the low-magnification ADF image of Fig.\ref{fig_additional_TEM}d highlights the possible growth mechanism leading to the hollow WS$_2$ pyramids. In this figure it can be observed that both layer-by-layer stacking and screw-dislocation-driven growth mechanisms contribute to the overall growth mechanism of the hollow WS$_2$ pyramids.}

\clearpage

\bibliographystyle{unsrt}

\bibliography{article_pyramids}

\end{document}